\documentclass[trackchanges]{aastex701}
\usepackage[utf8]{inputenc} 
\usepackage[T1]{fontenc}  
\usepackage{lmodern}

\begin{document}

\title{Earth-Mass Planets in Tandem Disks}

\author{Tokuhiro Nimura}
\affiliation{Okayama Astronomical Museum, 3037-5, Honjo, Kamogata, Asakuchi, Okayama 719-0232, Japan}
\email{nimura@oam-chikurinji.jp}  

\author{Toshikazu Ebisuzaki} 
\affiliation{RIKEN Center for Advanced Photonics, Photonics Control Technology Team, 2-1, Hirosawa, Wako, Saitama 351-0198, Japan}
\email{ebisu@mail.jmlab.jp}

\begin{abstract}

This paper presents a new terrestrial planet formation theory demonstrating that Earth-mass 
planets form naturally in tandem protosolar disks. Our model builds upon tandem planet formation 
theory \citep{Ebisuzaki2017,Imaeda2017a,Imaeda2017b,Imaeda2018}, 
incorporating magneto-rotational instability (MRI) suppression \citep{Balbus1991,Hawley1991}, 
porous particle aggregation \citep{Okuzumi2012,Kataoka2013} and standard planet formation mechanisms (e.g., \citealt{Safronov1969}; \citealt{Hayashi1985}). 
In tandem proto-solar disk, planets form at two distinct locations: the inner and outer edges of the MRI-suppressed region, 
where solid particles accumulate. The inner edge produces rocky planets, while the outer edge forms gas giants. 
When planetesimal reach Earth-sized mass at the inner MRI edge, they migrate outward due to gas disk torque. 
For a protosolar disk accretion rate of $\dot{M} = 10^{-7.08} M_\odot\, \mathrm{yr}^{-1}$ (Case D), 
the total solid mass at the inner MRI edge reaches $1.99 M_\oplus$, producing two Earth-mass planets. 
This result closely matches the distribution of terrestrial planets in the Solar System (Earth and Venus), 
which comprises of 92\% of total terrestrial planets mass, providing strong support for our formation mechanism.

\end{abstract}


\section{INTRODUCTION} 
Planets form in protoplanetary disk around young stars through gravitational 
collapse of dense molecular cloud cores (e.g., \citealt{Bouvier2007}). Solid 
particles grew from micron-sized dust grains of $\sim 10^{-10}$~g to 
planets of $\sim 10^{26}$--$10^{30}$~g.

In classical planetary formation scenarios (e.g., \citealt{Safronov1969}; \citealt{Hayashi1985}), 
planets were assumed to grow in individual orbits without radial migration. However, 
two major issues emerged. First, planetary growth rates were too slow to be consistent 
with solar system timescales. Specifically, Jupiter could not form within 4.6 billion 
years (e.g., \citealt{Wetherill1989}; \citealt{Ida1992a,Ida1992b,Ida1993}). Second, 
explaining the mass gap between rocky planets and gas giants, i.e., 1.5--5.0~AU (astronomical units) 
proved difficult (\citealt{Weidenschilling1977a}).

To address these issues, \citet{Walsh2011} proposed the Grand tack model 
with the following assumptions: (1) Jupiter first formed at 5~AU and migrated 
inward, concentrating solid particles near 1~AU. (2) Saturn formed and followed 
Jupiter 2:3 resonance. (3) Jupiter and Saturn locked became in resonance and 
migrated outward by 5--7~AU. The Grand tack migration concentrated solid 
particles around 0.7--1~AU. The high particle concentration accelerated 
rocky planet growth sufficiently to match with solar system observations and 
explained the 1.5--5.0~AU mass gap. Despite criticisms particularly questions 
about gas giants formation in outer orbits where planetary growth rates are typically 
low, this demonstrates that models concentrating solid particles near 1~AU can 
reproduce solar system like planetary architectures.

\textbf{Recent advances in planet formation theory have emphasized the role of pebble accretion, 
in which planetary cores grow efficiently by accreting small, drifting pebbles within protoplanetary disks. 
This process provides an efficient alternative to classical planetesimal accretion and offers a promising 
pathway for the rapid formation of giant planets (\citealt{Ormel2010,Johansen2017,Drazkowska2023,Ormel2024}). 
However, a sufficiently large and continuous supply of pebbles is essential for sustaining core growth (\citealt{Ikoma2025}). 
In addition, \citet{Bae2023} demonstrated that pressure bumps in gas disks can trap solids and strongly influence planet formation, 
although the physical origin of such structures remains uncertain, particularly in regions where planets have not yet formed.}

\textbf{Thus, while recent studies have highlighted the significance of pebble accretion and disk pressure structures, 
a unified understanding that consistently accounts for pebble supply, disk evolution, and core formation is still lacking. 
Therefore, a theoretical framework that naturally links these processes and explains the overall progression of planet formation is required.}

The tandem planet formation theory (\citealt{Ebisuzaki2017}; \citealt{Imaeda2017a,Imaeda2017b,Imaeda2018}) 
provides an alternative approach. Based on steady-state, 1-D accretion disk 
models incorporating magneto-rotational instability (MRI) 
(\citealt{Balbus1991}; \citealt{Hawley1991}), porous aggregation 
(\citealt{Okuzumi2012}; \citealt{Kataoka2013}) and radial particle drift 
(e.g., \citealt{Whipple1972}; \citealt{Adachi1976}; \citealt{Weidenschilling1977b}), 
these disks contain three regions (Figure 1): the inner turbulent region (ITR), 
the MRI-suppressed region (MSR), and the outer turbulent region (OTR). In ITR, 
the magneto-rotational instability occurs because the Elsasser number ($\Lambda$) 
exceeds unity (Figure 2b) due to thermal ionization of alkali metals (Na and K) at 
temperature ($T_\mathrm{m}$) above $1000~\mathrm{K}$ (Figure 2a). In the MSR, MRI is 
suppressed because the Elsasser number is lower than unity due to negligible 
ionization owing to the lower temperature than $1000~\mathrm{K}$. The MSR 
separates vertically into a quiet area (QA) and turbulent envelopes (Figure 1), 
with higher column density ($\Sigma$) than turbulent regions (ITR and 
OTR; Figure 2d). In the OTR, MRI resumes because the Elsasser number is larger 
than unity due to cosmic-ray ionization at a lower densities.

\begin{figure}[htbp]
 \centering
 \includegraphics[width=0.8\linewidth]{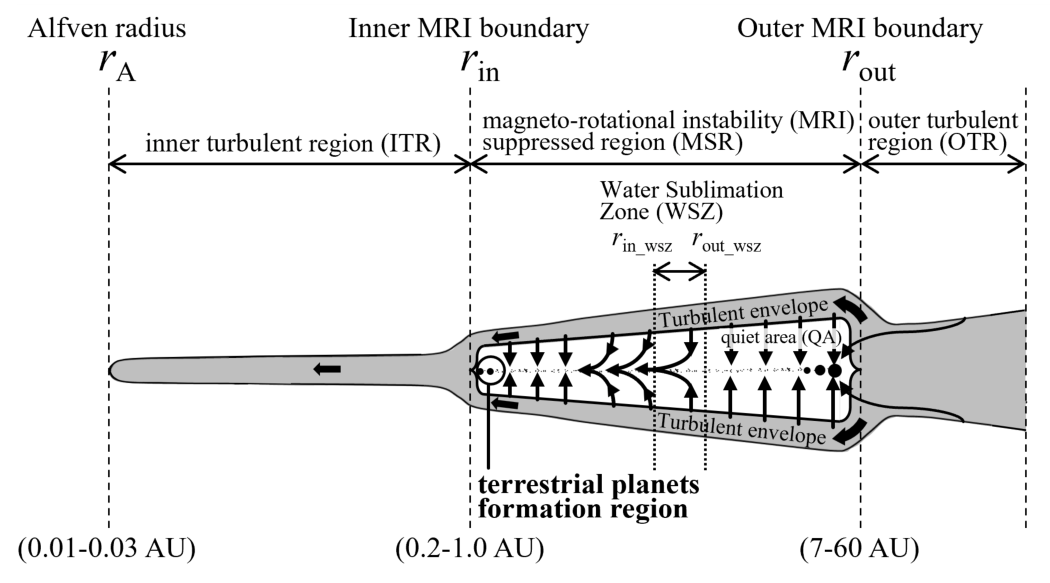}
 \caption{The structure of Tandem protoplanetary disk. The disk consists of three regions, 
the inner turbulent region (ITR), the magneto-rotational instability (MRI) suppressed region (MSR), 
and the outer turbulent region (OTR) (\citealt{Ebisuzaki2017}; \citealt{Imaeda2017a,Imaeda2017b,Imaeda2018}) }
 \label{fig:figure01}
\end{figure}
\begin{figure}[htbp]
 \centering
 \includegraphics[width=0.8\linewidth]{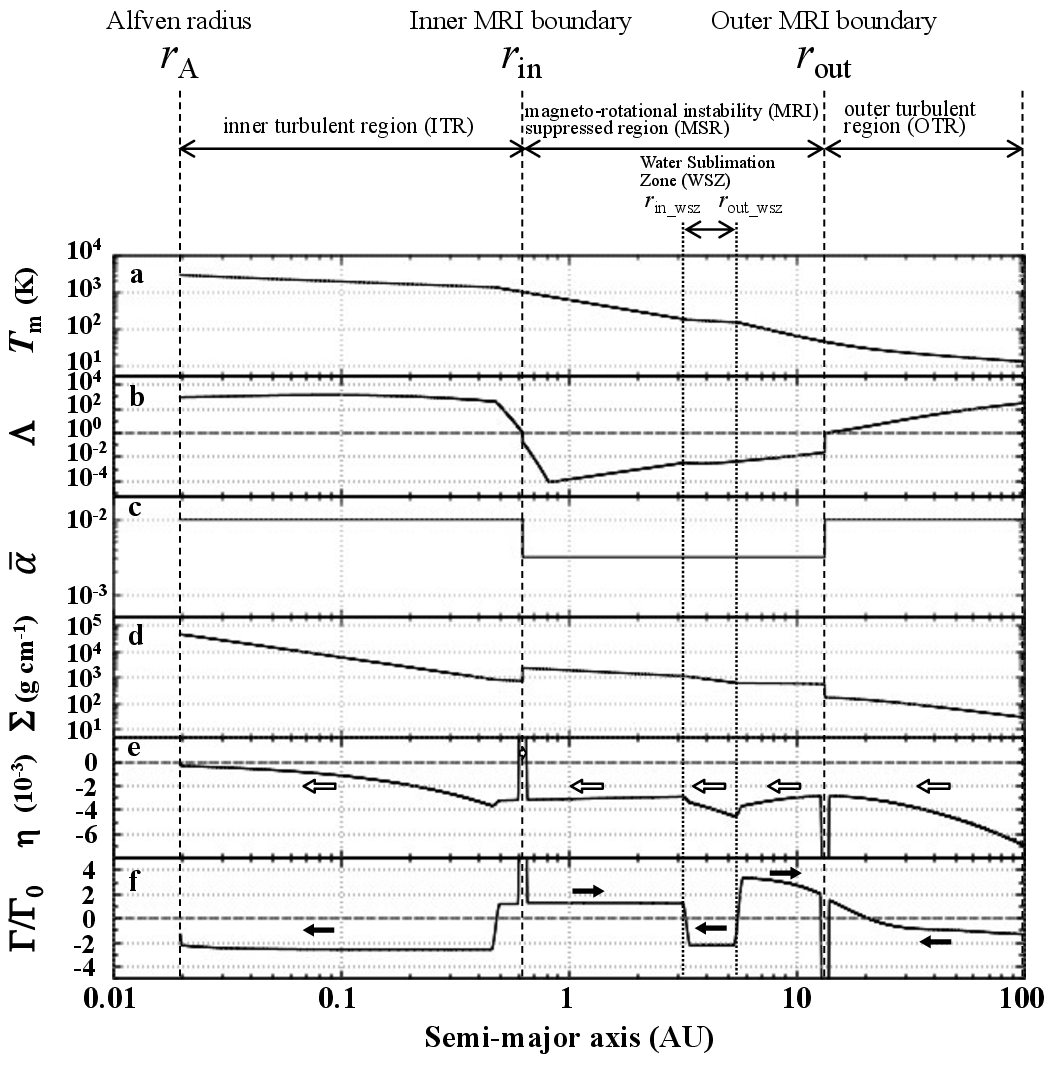}
 \caption{Radial profile of midplane temperature ($T_m$), Elsasser number ($\Lambda$), $\alpha$-value 
($\bar{\alpha}$), column density ($\Sigma$), $\eta$-value ($\eta$), and normalized torque ($\Gamma/\Gamma_0$) 
\citep{Paardekooper2009, Lyra2010, Paardekooper2010, Paardekooper2014} in the disk for accretion rate of 
$\dot{M} = 10^{-7.0} M_\odot\,\mathrm{yr}^{-1}$.}
 \label{fig:figure02}
\end{figure}

Rocky and icy planets form at the two boundaries, inner (ITR-MSR) and 
outer (MSR-OTR) where drifting solid particles accumulate. \citet{Ebisuzaki2017}, 
termed this ``Tandem planet formation theory'', because both regions function 
simultaneously as a planet factory. 
This theory naturally concentrates solid particles near 1~AU, 
reproducing \textbf{conditions required by both the Grand Tack and pebble accretion models, 
and can also account for the effects of pressure bumps in gas disks as demonstrated by \citet{Bae2023}}.

In tandem planetary formation theory, rocky planets are formed in the inner 
MRI front, as described below. First, micron-sized solid particles grow through 
mutual collisions in the QA of the MSR. When their sizes become as large as 
centimeters, they start to decuple with the gas and settle down on the equatorial 
plane to form a thin subdisk in the gas disk (Figure 1). We call it as 
``pebbles subdisk'' \citep{Cuzzi1993}. 
\textbf{In a protoplanetary disk, gas would orbit at the Keplerian velocity if gravitational forces were the only interactions. 
However, the presence of a radial pressure gradient modifies the force balance. In regions where the pressure decreases outward, 
the gas experiences an additional outward force, which reduces the rotational velocity compared with the Keplerian velocity 
because of the partial cancellation of the gravitational force due to this additional outward force.
Since the orbital velocity of solid particles are exactly that of the Keplerrian velocity, The solid particles orbits with slightly higher than that of gas.}

Therefore, the solid particles lose their angular momentum through gas 
drag, such that they drift toward the central star and eventually arrive at 
the inner MRI front. Since in the inner MRI front ($r_\mathrm{in}$), however, 
pressure gradient became minus because of the density gap (Figure 2e), 
and the inward drift motion of the particles vanished. The accumulated solid 
particles in the inner MRI front sank rapidly to the equatorial plane, and the 
density of the solid particles in the disk increased. Eventually, the gravitational 
instability takes place in the pebbles subdisk to form planetesimals. After that 
the rocky planetesimals rapidly grow in the inner MRI front swallowing continuously 
pebbles supplied from MSR.

The total mass of the rocky particles, $M_\mathrm{Rocky}$, originally located 
in the region between the inner MRI front ($r_\mathrm{in}$) and the inner edge of 
the water sublimation zone (WSZ) ($r_{\mathrm{in\_wsz}}$), is an important quantity 
in the tandem planet formation theory, as it is eventually the total mass of the 
rocky planets. It is estimated as $1$--$10\,M_{\oplus}$ (Earth masses), 
depending on the model parameters. This is promising since the total mass of 
rocky planets of solar system is $1.98\,M_{\oplus}$ (Mercury, Venus, Earth, and Mars) 
which is well within the range suggested by model above. The next natural 
question is how many planets are formed from this total mass and, equivalently, 
how large are the individual masses of these planets?

To answer these questions, we investigated the effect of the gravitational torque from the gas disk, 
which forces planets to migrate out of the inner MRI front when the planet\textquotesingle s mass becomes sufficiently high. 
\textbf{According to Equation (10) in \citet{Paardekooper2014}, the migration timescale of a planet is given by:}
\begin{eqnarray}
\tau_{\mathrm{mig}} = \frac{\pi}{2} \frac{h^2}{q_{\mathrm{d}}, q} \, \Omega_{\mathrm{planet}}^{-1}
\label{eq:migration_time_scale_Pa14}
\end{eqnarray}
\textbf{where, $h = H / r_{\mathrm{planet}}, q = M_{\mathrm{planet}} / M_*$ \citep{Paardekooper2014}.
Here $H$, $r_{\mathrm{planet}}$ and $M_{\mathrm{planet}}$ denote the vertical scale height of the gas disk, the semi-major axis 
and mass of the planet, respectively.
$\Omega_{\mathrm{planet}}$ is the Keplerian orbital frequency of the planet.
As shown in Equation (4) in \citet{Paardekooper2014}, the $q_{\mathrm{d}}$ is given as:}
\begin{eqnarray}
q_{\mathrm{d}} = \frac{\pi r_{\mathrm{planet}}^2 \, \Sigma_{\mathrm{planet}}}{M_*}
\label{eq:q_d_definition}
\end{eqnarray}
\textbf{From Equation (7) in Ebisuzaki and Imaeda (2017),  the vertical scale height of the gas disk ($H$) is given as:}
\begin{eqnarray}
H = \frac{c_{{\mathrm{s}},\mathrm{planet}}}{\Omega_{\mathrm{planet}}} 
= \sqrt{\frac{k_{\mathrm{B}} T_{\mathrm{m}} r_{\mathrm{planet}}^3}{\mu m_{\mathrm{H}} G M_*}}
\label{eq:H_definition}
\end{eqnarray}
\textbf{Here, $T_{\mathrm{m}}$, and $M_*$ denote the mid-plane temperature 
and the mass of central star, respectively.  $c_{\mathrm{s, planet}}$ is the isothermal sound velocity}
\textbf{$k_{\mathrm{B}}$, $\mu = 2.34$, $m_{\mathrm{H}}$ and $G$ are the Boltzmann constant, the mean molecular weight of gas, 
the mass of a hydrogen atom, and the gravitational constant respectively.}
\textbf{The $\Sigma_{\mathrm{planet}}$ is given as:}
\begin{eqnarray}
\Sigma_{\mathrm{planet}} = \frac{\dot{M} \, \Omega_{\mathrm{planet}}}{3 \pi \, c_{\mathrm{s,planet}}^2 \, \bar{\alpha}}
\label{eq:Sigma_planet}
\end{eqnarray}
\textbf{Here, $\dot{M}$ is the accretion rate of the protosolar disk. 
$\bar{\alpha}$ is the $\alpha$-value in the $\alpha$ disk assumption \citep{Shakura1973}. 
In the MRI suppressed region, it takes the value $\gamma \alpha_{\mathrm{act}}$, where $\alpha_{\mathrm{act}} = 1.0 \times 10^{-2}$ 
is the $\alpha$-value in the MRI active region \citep{Davis2010,Shi2010}. 
$\gamma = 10^{-0.5}$ is the reduction factor of turbulence due to the suppression 
of MRI \citep{Ebisuzaki2017}.
From Equation~(\ref{eq:migration_time_scale_Pa14}) to (\ref{eq:Sigma_planet}),}
\begin{eqnarray}
\tau_{\mathrm{mig}}&=& (0.78~\mathrm{Myr}) \left(\frac{\dot{M}}{10^{-7} M_\odot~\mathrm{yr}^{-1}}\right)^{-1} 
\left(\frac{T_{\mathrm{m}}}{1000~\mathrm{K}}\right)^2 \left(\frac{r_{\mathrm{planet}}}{\mathrm{AU}}\right)^2 
\left(\frac{M_{\mathrm{planet}}}{M_\oplus}\right)^{-1},
\label{eq:migration_time_scale}
\end{eqnarray}
Here, $M_\odot$ is the solar mass. As shown in Figure~3, the migration timescale ($\tau_{\mathrm{mig}}$) 
for an Earth-mass planet is as short as $0.78~\mathrm{Myr}$, which is shorter than the duration of accretion 
onto the central star (\(> 1$--$100~\mathrm{Myr}\)) (Figure~4). 
Therefore, in the present study, we investigated the effect of planet migration due to gravitational torque 
from a gas disk. As will be shown later, when a planet grows, 
it exits the planet-formation region (the MRI front). In Section~2, we describe 
the assumptions and methods. Section~3 presents the results of the calculations. 
Sections~4 and~5 present the discussion and conclusion, respectively.

\begin{figure}[htbp]
 \centering
 \includegraphics[width=0.8\linewidth]{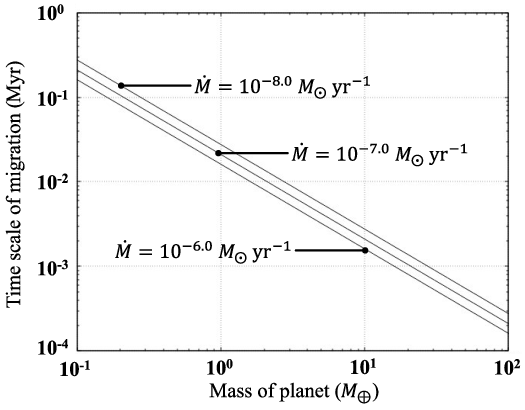}
 \caption{Time scale of migration and mass of planet. Each solid line means different of accretion rate.}
 \label{fig:figure03}
\end{figure}

\begin{figure}[htbp]
 \centering
 \includegraphics[width=0.8\linewidth]{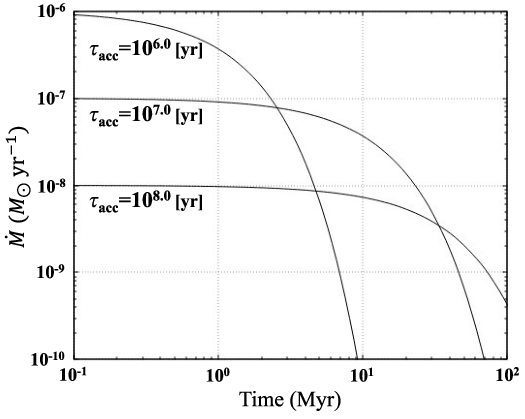}
 \caption{The accretion rate ($\dot{M}$) is assumed to decrease exponentially with a decay timescale, with the central star mass set to $1\,M_\odot$.}
 \label{fig:figure04}
\end{figure}

\section{ASSUMPTIONS AND METHOD} 
We determine the accretion disk structure around a $1\,M_\odot$ central star 
and calculate planetary growth following \citet{Ebisuzaki2017} with two modifications. 
First, rather than assuming constant accretion rates, we adopt exponentially decaying accretion:
\begin{equation}
\dot{M}(t) = \dot{M}_0 \exp\left(-\frac{t}{\tau_{\mathrm{acc}}}\right),
\label{eq:mdot_decay}
\end{equation}
where $t$ is the time, and $\dot{M}_0$ and $\tau_{\mathrm{acc}}$ ($10^5$–$10^8$ yr) are 
the initial accretion rate and the decay constant of the accretion rate, respectively. 
Here, we set $\tau_{\mathrm{acc}}$ as:
\begin{equation}
\tau_{\mathrm{acc}} = \frac{1\,M_\odot}{\dot{M}_0} \,\mathrm{yr},
\label{eq:tau_acc}
\end{equation}
so that the total accreted mass is $1 M_\odot$, compatible with the mass of the central star. 
Second, we assume the equilibrium particle scale height, $H_{\mathrm{pe},k}$, of the Lagrangian 
superparticle $k$ follows:
\begin{equation}
H_{\mathrm{pe},k} = H_k 
\left(1 + \frac{\Omega_k t_{\mathrm{s},k}}{\alpha_{\mathrm{D},k}} \right)^{-1/2} 
\left(1 + \frac{\Omega_k t_{\mathrm{s},k}}{\Omega_k t_{\mathrm{s},k} + 1} \right)^{-1/2}
+ \frac{e_{\mathrm{VS},k}}{2}, 
\label{eq:scale_height}
\end{equation}
where $H_k$, $\Omega_k$, $t_{\mathrm{s},k}$, and $\alpha_{\mathrm{D},k}$ are the vertical scale height of the gas disk, 
the Keplerian orbital frequency, particle stopping time, and viscous parameter for the Lagrangian 
superparticle $k$, respectively. Unlike \cite{Ebisuzaki2017}, we include 
planetary perturbation effects via the final term in Equation~(\ref{eq:scale_height}). 
The equilibrium eccentricity, $e_{\mathrm{VS},k}$, results from competition 
between planetary stirring and gas drag damping, governed by:
\begin{equation}
e_{\mathrm{VS},k}^5 + \frac{\pi \eta_k}{3.422} e_{\mathrm{VS},k}^4 
- \frac{1}{3.422} \frac{6 m_{\mathrm{p},k}}{2^{1/3} \pi C_{\mathrm{D},k} a_{\mathrm{p},k}^2 \rho_{\mathrm{m}} r_{\mathrm{p},k}} \left(\frac{m}{3 M_*}\right)^{5/3} = 0,
\label{eq:eccentricity}
\end{equation}
\citep{Adachi1976,Youdin2007}, where $a_{\mathrm{p},k}$ is the particle radius, 
$\rho_{\mathrm{m}}$ is the gas density at the midplane, $r_{\mathrm{p},k}$ (or $r$) is the distance from 
the central star, and $C_{\mathrm{D},k}$ is the aerodynamic coefficient \citep{Ebisuzaki2017}. 
The parameter $\eta_k$ follows the definition given by \citet{Ebisuzaki2017}, and 
its behavior is illustrated in Figure~2e. $M_*$ is the mass of the central star. 
Here, $m$ is set to be the mass of the pebble ($m_{\mathrm{p},k}$) before a planet is formed, 
and the mass of the planet ($M_{\mathrm{planet}}$) after a planet is formed.

In this study, we calculated planetary growth in the MRI front as follows. The growth rate is:
\begin{eqnarray}
\frac{dM_{\mathrm{planet}}}{dt} 
&=& \frac{1}{1.3 \times 10^5} 
\left( \frac{\Sigma}{2400~\mathrm{g~cm}^{-2}} \right)^{2/5}
\left( \frac{M_{\mathrm{planet}}}{M_\oplus} \right)^{2/3}
\left( \frac{\rho_{\mathrm{i},k}}{2~\mathrm{g~cm}^{-3}} \right)^{-3/5} 
\nonumber \\ 
&& \times
\left( \frac{r_{\mathrm{planet}}}{\mathrm{AU}} \right)^{-3/5}
\left( \frac{\Sigma_{\mathrm{p}}}{10~\mathrm{g~cm}^{-2}} \right)
\left( \frac{m_{\mathrm{p}}}{10^{18}~\mathrm{g}} \right)^{-2/15} 
\frac{M_\oplus}{\mathrm{yr}},
\label{eq:planetary_growth_rate}
\end{eqnarray}
\citep{Kokubo2012}, where $\rho_{\mathrm{i},k} = 2~\mathrm{g~cm}^{-3}$ is the density of the solid particle. 
The $\Sigma_{\mathrm{p}}$ and $m_p$ are the column density and the mass of a solid particle, respectively. 
In contrast, \citet{Ebisuzaki2017} assumed that the mass of the planet was equal to the total mass of the solid particles in the MRI front.

Next, as the planet grows, the gravitational torque exerted on the planet by the gas disk becomes 
significant and the planet migrates from the MRI front, where the pebbles are concentrated.
The velocity of the migration of the planet, $v_{r,\mathrm{planet}}$, owing to the gravitational 
torque from the gas disk is calculated as:

\begin{equation}
v_{r,\mathrm{planet}} = \frac{dr_{\mathrm{planet}}}{dt} = \frac{2 \Gamma}{M_{\mathrm{planet}} r_{\mathrm{planet}} \Omega_{\mathrm{planet}}},
\label{eq:v_r_planet}
\end{equation}

\citep{Paardekooper2009, Lyra2010, Paardekooper2010, Paardekooper2014}. 
Gravitational torque, $\Gamma$, is calculated in terms of the logarithmic gradients, 
$\alpha\, (= -\partial \ln \Sigma / \partial \ln r)$ and 
$\beta\, (= -\partial \ln T_{\mathrm{m}} / \partial \ln r)$, as:
\begin{eqnarray*}
&& \hspace*{3.8cm} 
\Gamma = \frac{\bigl( -0.61 - 2.33\alpha + 2.82\beta \bigr) \Theta^2 + \bigl( -0.85 - \alpha - 0.9\beta \bigr)}{(\Theta + 1)^2} \, \Gamma_0, 
\hspace*{5.15cm} \makebox[0pt][r]{(8a)} \\[2ex]
&& \hspace*{3.8cm} 
\Theta = \frac{ c_{\mathrm{V}} \Sigma \Omega_{\mathrm{planet}}}{ 12\pi\sigma T_{\mathrm{m}}^3 } \left( \frac{3\tau}{8} + \frac{\sqrt{3}}{4} + \frac{1}{4\tau} \right), 
\hspace*{8.46cm} \makebox[0pt][r]{(8b)}
\end{eqnarray*}
\setcounter{equation}{8} 
\citep{Paardekooper2009,Lyra2010,Paardekooper2010,Paardekooper2014}, 
where $ c_{\mathrm{V}} $, $\sigma$, and $\tau$ are the isochoric heat capacity, 
the Stefan–Boltzmann constant, and the optical depth of the disk, respectively. 
\textbf{In this paper, the symbol $\alpha$ is employed in two distinct contexts: 
firstly, it denotes the Shakura--Sunyaev viscosity parameter; 
secondly, it represents the logarithmic surface--density gradient, 
$\alpha = -\partial \ln \Sigma / \partial \ln r$. 
This follows the conventional notation utilised in the extant literature, 
and the same symbol is retained here. 
The distinction between the two definitions is determined by the respective 
contexts in which they are employed.}
The torques were normalized using:
\begin{equation}
\Gamma_0 =
\frac{q^2}{h^2} \, \Sigma \, r_{\mathrm{planet}}^4 \, \Omega_{\mathrm{planet}}^2,
\label{eq:gamma_0}
\end{equation}
\citep{Paardekooper2009,Lyra2010,Paardekooper2010,Paardekooper2014}. Here, a planet migrates outward when 
$\Gamma$ is positive, and vice versa (Figure~2f).

When two planets approach each other, we keep the smaller planet away from 
the larger planet at a distance larger than five times the Hill radius:
\begin{equation}
\Delta R = 5 \left(\frac{M_{\mathrm{planet}}}{3M_*}\right)^{1/3} r_{\mathrm{planet}},
\label{eq:delta_R}
\end{equation}
\citep{Kokubo1995}, considering the repulsive force caused by mutual scattering.

We calculated five cases (A, B, C, D, and E), as shown in Figure~5 and Table~1. 
Case A adopts $\dot{M}_0=10^{-7.00} \; M_\odot\,\mathrm{yr}^{-1}$ and $f_{\mathrm{p}}=2.5\times10^{-3}$ 
(particle fraction) following \cite{Ebisuzaki2017}. Cases B, C, and D, adjust $\dot{M}_0$ 
to achieve $M_{\mathrm{Rocky}} \sim 1.98M_\oplus$, matching the total planet mass 
in our solar system (the sum of Mercury, Venus, Earth, and Mars; dashed line in Figure~5). 
In cases A, B, and E, we adopt $f_{\mathrm{p}} = 2.5 \times 10^{-3}$, which is assumed by 
\citet{Ebisuzaki2017}. However, $f_{\mathrm{p}}$ is not adequately constrained. In fact, it has a large 
value range of $0.001$--$0.1$ as estimated by the Atacama Large Millimeter/Submillimeter 
Array (ALMA) Survey \citep{Ansdell2016}. Therefore, in cases C and D, $f_{\mathrm{p}}$ was assumed 
to be $2$ and $0.5$, respectively, as in \citet{Ebisuzaki2017}. In case E, we adopt 
$M_{\mathrm{Rocky}} = 10.0 M_\oplus$ and $f_{\mathrm{p}} = 2.5 \times 10^{-3}$, 
when $\dot{M} = 10^{-6.75}\ M_\odot\ \mathrm{yr}^{-1}$.

\begin{deluxetable*}{cccccccccccc}
\tablecaption{Disk parameters, formed planets, and mass accretion parameters.}
\tablehead{
\colhead{} & \multicolumn{3}{c}{Disk parameters} & \multicolumn{3}{c}{Formed planets} & \multicolumn{4}{c}{Mass accretion parameters Equation (\ref{eq:M_MRI})} \\
\hline
\colhead{Case} & 
\colhead{$\dot{M}$} & 
\colhead{$f_{\mathrm{p}}$} & 
\colhead{$\tau_{\rm acc}$} & 
\colhead{Planet \#} & 
\colhead{$r_{\rm planet}$} &
\colhead{$M_{\rm planet}$} & 
\colhead{$t_{\rm S,acc}$} & 
\colhead{$t_{\rm E,acc}$} & 
\colhead{$M_{\rm MRI,0}$} & 
\colhead{$M_{\rm Rocky}$} \\
\colhead{} &
\colhead{($M_\odot~{\rm yr}^{-1}$)} & 
\colhead{($10^{-3}$)} & 
\colhead{(yr)} & 
\colhead{} & 
\multicolumn{1}{c}{(AU)} &
\colhead{($M_\oplus$)} & 
\colhead{($10^{-3}$ Myr)} & 
\colhead{($10^{-2}$ Myr)} & 
\colhead{($10^{-2}~M_\oplus$)} & 
\colhead{($M_\oplus$)}
}

\startdata
A & $10^{-7.00}$ & 2.50$^\ast$ & $10^{7.00}$ & 1 & 3.18 & 2.18 & 4.31 & 3.93 & 9.65 & 4.92 \\
  &              &             &             & 2 & 2.79 & 2.26 &      &      &      &      \\
  &              &             &             & 3 & 2.50 & 0.46 &      &      &      &      \\
B & $10^{-7.33}$ & 2.50$^\ast$ & $10^{7.33}$ & 1 & 2.25 & 1.78 & 2.49 & 2.38 & 3.71 & 1.97 \\
  &              &             &             & 2 & 2.05 & 0.193&      &      &      &      \\
C & $10^{-7.58}$ & 5.00        & $10^{7.58}$ & 1 & 1.82 & 1.96 & 0.927& 0.939& 3.59 & 1.96 \\
D & $10^{-7.08}$ & 1.25        & $10^{7.08}$ & 1 & 1.54 & 0.974& 6.76  & 6.17 & 3.83 & 1.99 \\
  &              &             &             & 2 & 1.39 & 1.04 &      &      &      &      \\
  &              &             &             & 3 & 1.31 & 0.0115&     &      &      &      \\
E & $10^{-6.75}$ & 2.50$^\ast$ & $10^{6.75}$ & 1 & 4.17 & 2.51 & 6.12 & 6.05 & 19.8 & 10.0 \\
  &              &             &             & 2 & 3.64 & 2.64 &      &      &      &      \\
  &              &             &             & 3 & 3.17 & 2.62 &      &      &      &      \\
  &              &             &             & 4 & 2.77 & 2.24 &      &      &      &      \\
\enddata
\tablecomments{$^\ast$ Based on Ebisuzaki \& Imaeda (2017).}
\end{deluxetable*}

\begin{figure}[htbp]
 \centering
 \includegraphics[width=0.8\linewidth]{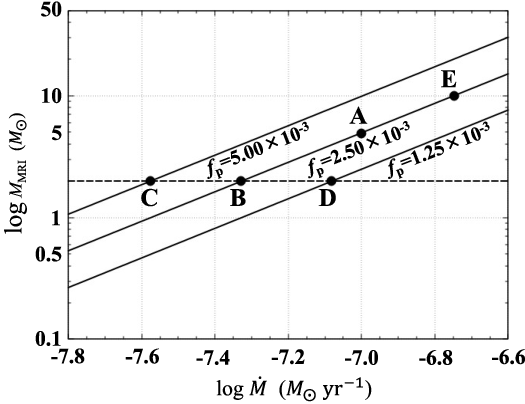}
 \caption{Mass of the terrestrial planet formation region. We assume that the mass is decided 
by accretion rate ($\dot{M}$) and particle fraction ($f_{\mathrm{p}}$)  of protosolar disk. 
A dash line is the total mass of the rocky planets of the solar system ($M_{\mathrm{Rocky}} \sim 1.98\,M_{\oplus}$)  
(the sum of Mercury, Venus, Earth, and Mars).}
 \label{fig:figure05}
\end{figure}

\begin{figure*}[htbp]
  \centering
   \begin{minipage}[b]{0.45\linewidth}
    \includegraphics[width=\linewidth]{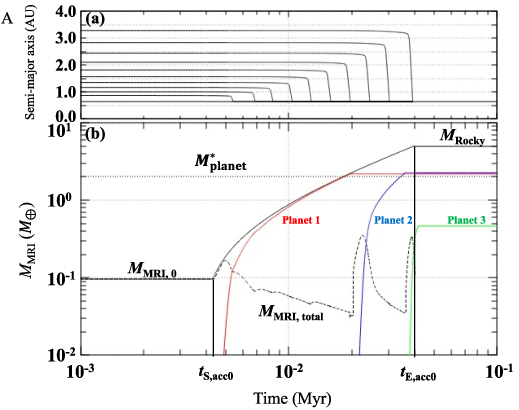}
  \end{minipage}
  \hspace{0.02\linewidth}
  \begin{minipage}[b]{0.45\linewidth}
    \includegraphics[width=\linewidth]{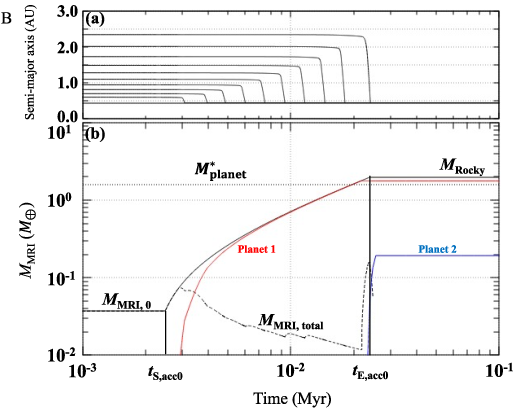}
  \end{minipage}\\[1ex]

  \begin{minipage}[b]{0.45\linewidth}
    \includegraphics[width=\linewidth]{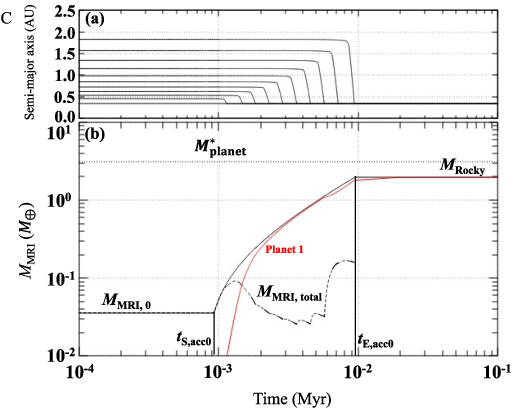}
  \end{minipage}
  \hspace{0.02\linewidth}
  \begin{minipage}[b]{0.45\linewidth}
    \includegraphics[width=\linewidth]{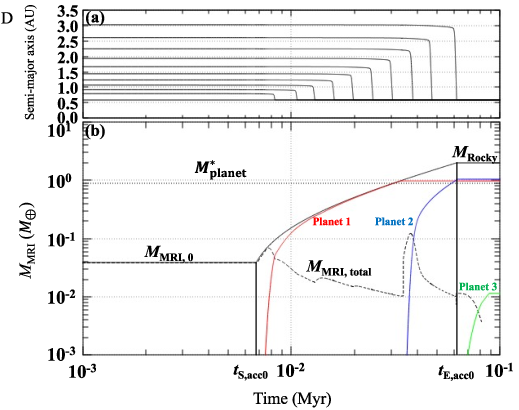}
  \end{minipage}\\[1ex]

  \hspace{0.03\linewidth}%
  \begin{minipage}[b]{0.45\linewidth}
    \includegraphics[width=\linewidth]{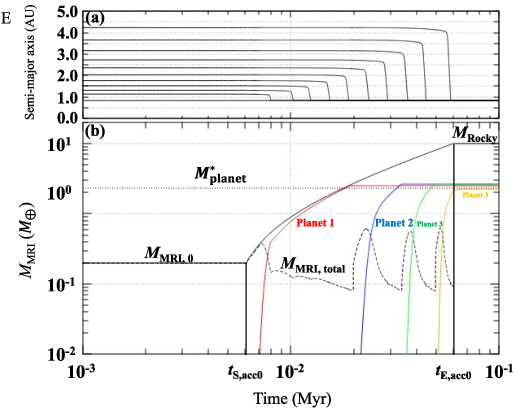}
  \end{minipage}
  \hfill
  \mbox{}

\caption{
Mass accretion to the inner MRI front from the terrestrial planet formation region. 
\textbf{(a) Semi-major axes of the superparticles accumulating at the inner MRI front.} 
\textbf{(b) Evolution of the MRI front mass ($M_{\mathrm{MRI}}$) due to accretion, as defined by Equation~(\ref{eq:M_MRI}), with relevant parameters listed in Table~1.} 
\textbf{Although the vertical axis indicates $M_{\mathrm{MRI}}$, other quantities, such as planetary masses, are also plotted for comparison.} 
Solid colored lines represent the mass of each planet. 
The dashed line shows the mass variation of solid particles in the inner MRI front, and the dotted line represents 
the analytical planetary mass given by Equation~(\ref{eq:planetmass}). 
The vertical dashed lines indicate the start and end times 
of the drift into the inner MRI front, denoted as $t_{\mathrm{S,acc}}$ and $t_{\mathrm{E,acc}}$, respectively.
}
  \label{fig:figure06}
\end{figure*}

\section{ROCKY PLANET FORMATION} 
Figure~6 shows the massive evolution ($M_{\mathrm{MRI}}$) of solid particles in the MRI 
front due to pebble drift, represented by Lagrangian superparticle in the five cases calculated in Table~1. 
The time evolution of $M_{\mathrm{MRI}}$ is well expressed by:
\begin{eqnarray}
M_{\mathrm{MRI}} &=& \left\{
\begin{array}{ll}
M_{\mathrm{MRI},0} & (t < t_{\mathrm{S,acc}}) \\
\frac{dM_{\mathrm{MRI,total}}}{dt} (t - t_{\mathrm{S,acc}}) + M_{\mathrm{MRI},0} & (t_{\mathrm{S,acc}} \leq t < t_{\mathrm{E,acc}}) \\
M_{\mathrm{Rocky}} & (t_{\mathrm{E,acc}} \leq t)
\end{array}
\right.,
\label{eq:M_MRI}
\end{eqnarray}
where $M_{\mathrm{MRI},0}$ is the initial mass at the inner MRI front. The start and end 
times of the drift into the inner MRI front are denoted as $t_{\mathrm{S,acc}}$ and 
$t_{\mathrm{E,acc}}$, respectively. These values are listed in Table~1.

In Case~A (Table~1; Figure~7A), orbit 2 merges into orbit 1 (the MRI front) at $t = 4.31 
\times 10^{-3}$ Myr, initiating $M_{\mathrm{MRI}}$ growth. As orbits 3--11 merge successively, 
$M_{\mathrm{MRI}}$ increases, reaching $M_{\mathrm{Rocky}} = 4.92 M_\oplus$ at $3.93 \times 10^{-2}$ 
Myr (Table~1; Figure~6A). This represents a 50-fold concentration enhancement at the MRI front.

\clearpage
\noindent
\begin{center}
\includegraphics[width=0.9\linewidth]{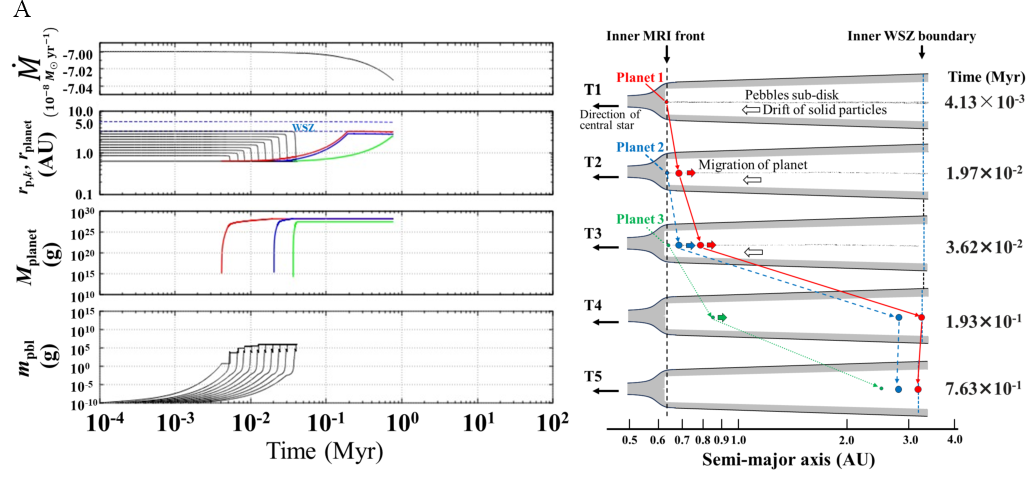}\\[1ex]
\includegraphics[width=0.9\linewidth]{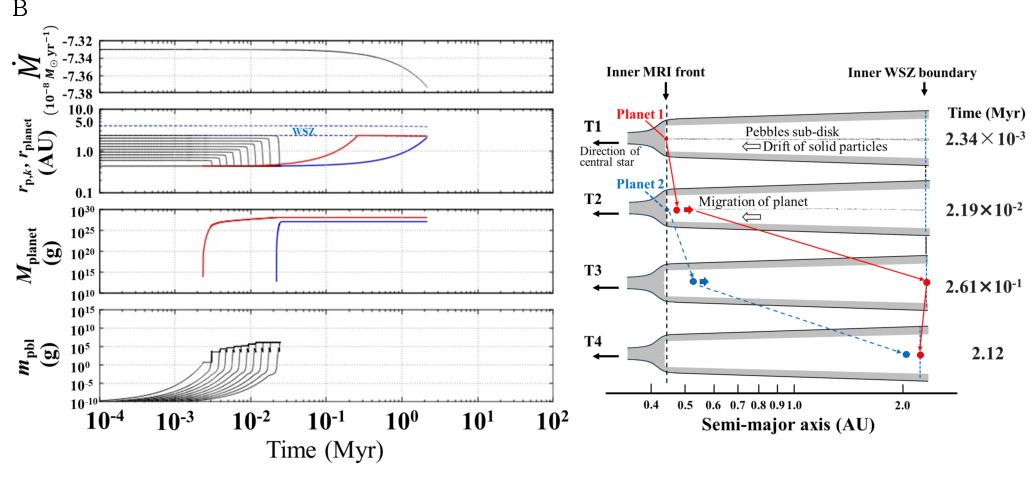}\\[1ex]
\includegraphics[width=0.9\linewidth]{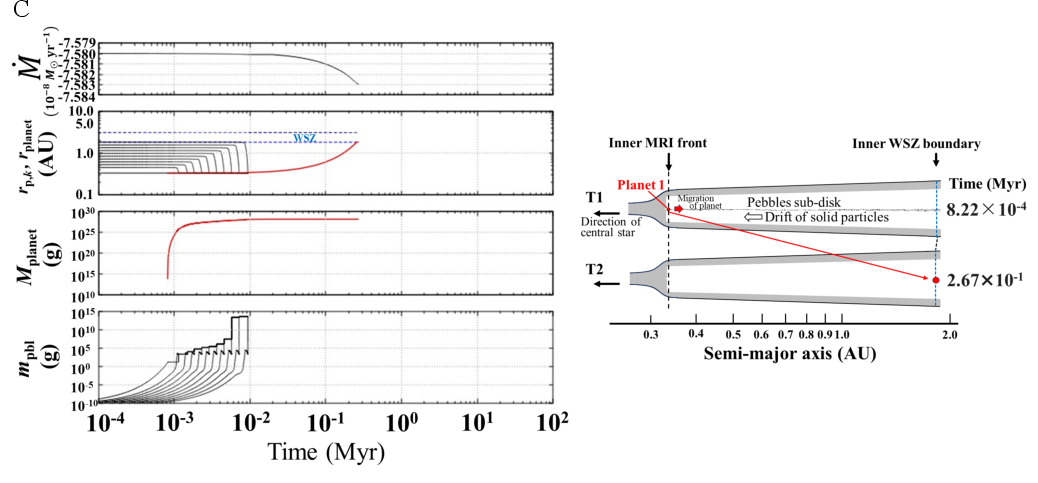}\\[1ex]
\includegraphics[width=0.9\linewidth]{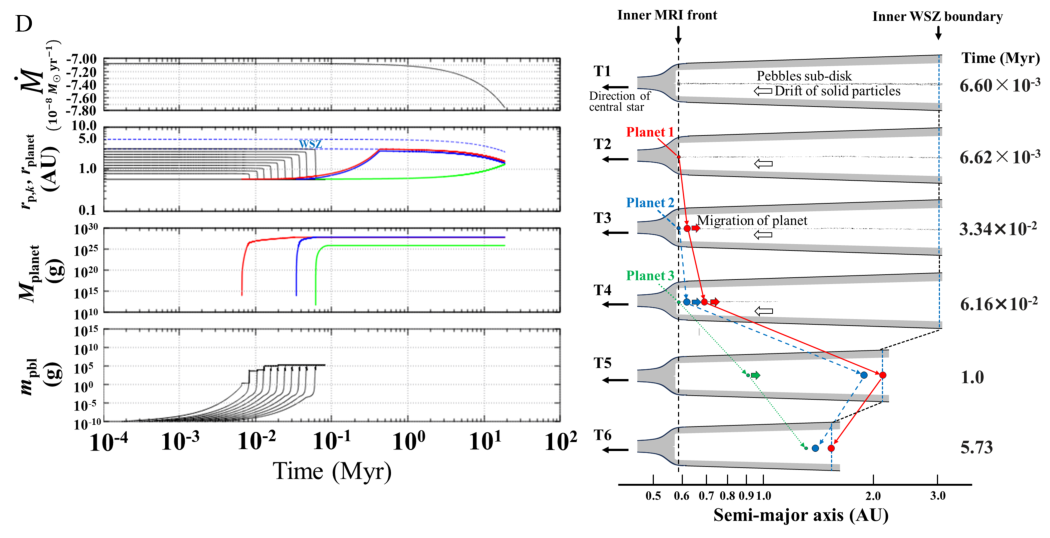}\\[1ex]
\includegraphics[width=0.9\linewidth]{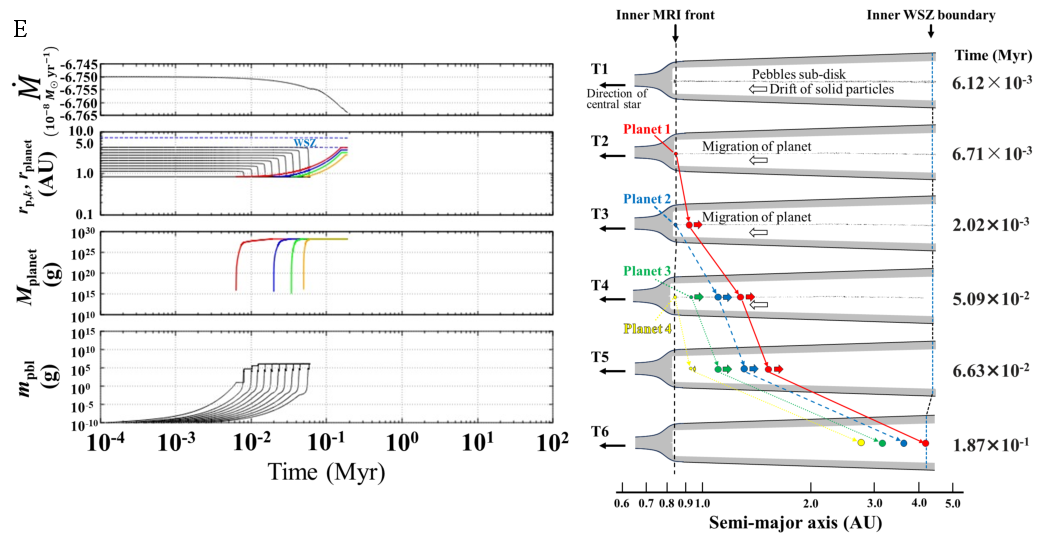}\\[2ex]
\end{center}
\addtocounter{figure}{1}

\noindent
\textbf{Figure 7.} Origin and evolution of terrestrial planets in Tandem protosolar disk. In the left graph, each panels show mass accretion ($\dot{M}$) 
, semi-major axis of pebbles and planets $(r_{\mathrm{p}}, r_{\mathrm{planet}})$, mass of planets $(M_{\mathrm{planet}})$, 
and mass of pebbles $(m_{\mathrm{pbl}})$. The solid colored line show orbital and mass evolution of planets. In right schematic illustration, planetary 
formation scenario of solar-like system. Each panel corresponds to a different stage of evolution. When the time scale of the orbital 
evolution ($\tau_{r_{\mathrm{planet}}}$) of all planets exceeds the time scale of disk evolution $(\tau_{\mathrm{disk}})$, the calculations are stopped.
\clearpage

\begin{figure}[htbp]
\centering
\includegraphics[width=0.8\linewidth]{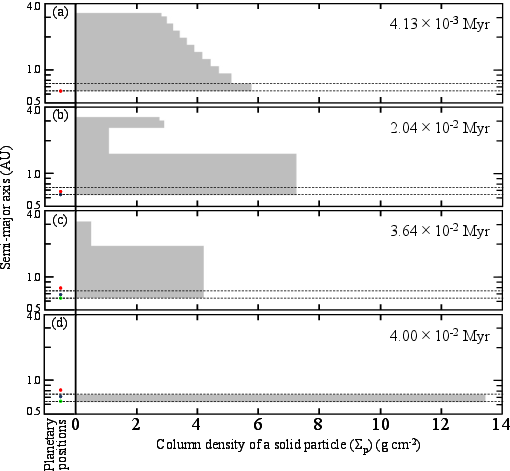}
\caption{\textbf{Evolution of the column density of solid particles, $\Sigma_{\mathrm{p}}$, for  $\dot{M} = 10^{-7.00} M_\odot\,\mathrm{yr}^{-1}$. 
Each panel shows the column density of solid particles at the formation time of each planet (Figure~7A) and at $4.00 \times 10^{-2}\,\mathrm{Myr}$.  
Solid particles drift and accumulate at the MRI front, resulting in an enhanced $\Sigma_{\mathrm{p}}$. 
After formation, planets migrate outward from the MRI front. Once the planets migrate outward, 
inward pebble accretion becomes inefficient compared to the planetary growth timescale (Figure~9), and planetary growth effectively ceases. 
Even after the planets pass through regions where $\Sigma_{\mathrm{p}}$ is zero, no further planetary growth occurs.}}
\label{fig:figure08}
\end{figure}

\begin{figure}[htbp]
\centering
\includegraphics[width=0.8\linewidth]{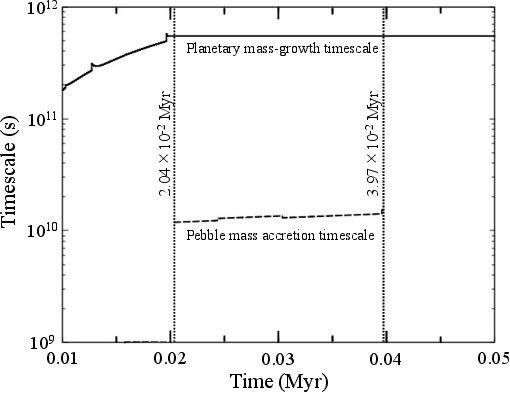}
\caption{\textbf{Planetary mass-growth timescale of the first-formed planet (Planet~1) and the pebble radial-drift timescale of pebbles surrounding Planet~1 
toward the MRI front for $\dot{M} = 10^{-7.00}\,M_\odot\,\mathrm{yr}^{-1}$. 
After Planet~1 migrates outward from the MRI front at $2.04 \times 10^{-2}\,\mathrm{Myr}$, it remains within the pebble flow for a limited time. 
However, the pebble radial-drift timescale toward the MRI front is shorter than the planetary growth timescale, preventing efficient pebble accretion. 
At $3.97 \times 10^{-2}\,\mathrm{Myr}$, Planet~1 exits the pebble flow.}}
\label{fig:figure09}
\end{figure}

\begin{deluxetable*}{ccccccccc} 
\tablecaption{Analytical calculation results of planetary mass using Equation~(\ref{eq:planetmass}) with simulation results.}
\tablehead{
\colhead{Case} & 
\colhead{$\dot{M}$} & 
\colhead{$f_p$} & 
\colhead{$\dot{M}_{\rm MRI,total}$} & 
\colhead{$\Sigma$} & 
\colhead{$M_{\rm planet}^*$} & 
\colhead{$M_{\rm Rocky}$} & 
\colhead{$M_{\rm Rocky}/M_{\rm planet}^*$} & 
\colhead{Simulation results} \\
\colhead{} &
\colhead{($M_\odot\,{\rm yr}^{-1}$)} & 
\colhead{($10^{-3}$)} & 
\colhead{($10^{16}\,{\rm g\,s}^{-1}$)} & 
\colhead{(${\rm g\,cm}^{-2}$)} & 
\colhead{($M_\oplus$)} & 
\colhead{($M_\oplus$)} & 
\colhead{} & 
\colhead{($M_\oplus$)}
}
\startdata
A & $10^{-7.00}$ & 2.50$^\ast$ & 2.61 & 2305 & 2.0 & 4.2 & 2.1 & \multicolumn{1}{c}{2.18} \\
  &              &             &      &      &     &     &     & \multicolumn{1}{c}{2.26} \\
  &              &             &      &      &     &     &     & \multicolumn{1}{c}{0.46} \\
B & $10^{-7.33}$ & 2.50$^\ast$ & 1.71 & 1815 & 1.6 & 1.6 & 1.0 & \multicolumn{1}{c}{1.78} \\
  &              &             &      &      &     &     &     & \multicolumn{1}{c}{0.193} \\
C & $10^{-7.58}$ & 5.00        & 4.3  & 1516 & 3.1 & 1.6 & 0.5 & \multicolumn{1}{c}{1.96} \\
D & $10^{-7.08}$ & 1.25        & 0.66 & 2174 & 0.9 & 1.7 & 1.9 & \multicolumn{1}{c}{0.974} \\
  &              &             &      &      &     &     &     & \multicolumn{1}{c}{1.04} \\
  &              &             &      &      &     &     &     & \multicolumn{1}{c}{0.0115} \\
E & $10^{-6.75}$ & 2.50$^\ast$ & 3.42 & 2766 & 2.3 & 8.4 & 3.6 & \multicolumn{1}{c}{2.51} \\
  &              &             &      &      &     &     &     & \multicolumn{1}{c}{2.64} \\
  &              &             &      &      &     &     &     & \multicolumn{1}{c}{2.62} \\
  &              &             &      &      &     &     &     & \multicolumn{1}{c}{2.24} \\
\enddata
\tablecomments{$^\ast$ Based on Ebisuzaki \& Imaeda (2017).}
\end{deluxetable*}

Increasing particle density triggers gravitational instability, forming planetesimal that grow 
into planet through pebble accretion. 
\textbf{We adopted the gravitational instability criterion for the dust layer derived by \citet{Yamoto2004}, 
which accounts for the competition between vertical shear (Kelvin–Helmholtz instability) and self-gravity in the midplane layer.}
Figure~7A shows the orbital and mass evolutions of 
Lagrangian superparticles and planets.

The first planet forms at $4.13 \times 10^{-3}$ Myr via gravitational instability (Figure~7A, T1) and 
grows through pebble accretion. When its mass reaches $2.18 M_\oplus$ at $1.97 \times 10^{-2}$ 
Myr, gravitational torque from the gas disk causes outward migration from the inner MRI front
 (Equations~(\ref{eq:v_r_planet})--(\ref{eq:gamma_0})), halting growth 
\textbf{as the planet moves beyond the region where solid particles are concentrated 
(approximately five Hill radii away, Equation~(\ref{eq:delta_R}))}, 
which prevents it from accreting all pebbles and allows subsequent planets to form
 (Figure~7A, T2). At this point, the planet\textquotesingle s distance exceeds $\Delta R$ 
(Equation~(\ref{eq:delta_R})) from the inner MRI front. The planet mass at that time is estimated as:
\begin{eqnarray}
M_{\mathrm{planet}}^* &\simeq& 0.9 \left[
\frac{
\ln(10) \left( 3 \log_{10} \left\{ f_{\mathrm{p}}/(1.25 \times 10^{-2}) + 8.4 \right\} \right) + (2 \times 3)/5
}{21}
\right]^{-3/5} 
\left(\frac{\Gamma}{1.21 \Gamma_0}\right)^{-3/5} 
\left(\frac{D}{5}\right)^{3/5} 
\nonumber \\
&& \times \left(\frac{T_{\mathrm{m}}}{1000\,\mathrm{K}}\right)^{1/30} 
\left(\frac{M_*}{M_\odot}\right)^{3/5} 
\left(\frac{\dot{M}}{10^{-7.08} M_\odot\, \mathrm{yr}^{-1}}\right)^{4/15} 
\left(\frac{f_{\mathrm{p}}}{1.25 \times 10^{-3}}\right)^{6/5} M_\oplus.
\label{eq:planetmass}
\end{eqnarray}

The derivation of Equation~(\ref{eq:planetmass}) and definitions of $m_{\mathrm{p}}^*$ and $m_0^*$ are provided in the 
Appendix. $D=5$ is the distance from the planet normalized by the Hill radius. $\beta_{\kappa} = 2.13 
\times 10^{-2}$ is a coefficient in the relationship $\kappa = \beta_{\kappa} T_{\mathrm{m}}^{0.75}$, applicable to 
the temperature range (180--1380 K) in regions of terrestrial planet formation \citep{Stepinski1998}.

The analytically estimated masses of cases A, B, C, D, and E are shown in Table~2 and are nearly equal 
to the mass obtained by simulation unless pebbles are used. The mass of terrestrial planets 
($M_{\mathrm{planet}}^*$) increased depending on the particle fraction ($f_{\mathrm{p}}$), mass of the central star 
($M_*$), and accretion rate ($\dot{M}$).

\textbf{Solid particles drift inward and accumulate at the MRI front, producing a localized enhancement of the solid particle column density, $\Sigma_{\mathrm{p}}$ (Figure~8). 
Planets initially form in this high-$\Sigma_{\mathrm{p}}$ region near the MRI front. After formation, the planets migrate outward away from the MRI front and leave the region of enhanced solid particle density. 
Figure~9 shows that once the first-formed planet (Planet~1) migrates outward from the MRI front at $3.44 \times 10^{-2}\,\mathrm{Myr}$, 
the radial drift timescale of pebbles surrounding Planet~1 toward the MRI front becomes shorter than the planetary mass-growth timescale.
Because pebble accretion is then inefficient, planetary growth effectively ceases. Even after the planet passes through regions 
where $\Sigma_{\mathrm{p}}$ is zero, no further planetary growth occurs.}

After the first planet migrates outward, second planet formation becomes possible as pebble subdisk 
thickness decreases due to reduced stirring. At $3.62 \times 10^{-2}$ Myr, the second planet becomes 
massive enough to migrate outward (Figure~7A, T3). A third planet then forms but remains smaller due to depleted pebble supply.

These planets migrate outward via gravitational torque (Equations~(\ref{eq:v_r_planet})--(\ref{eq:gamma_0})) and stabilize near the inner 
water sublimation zone (WSZ) boundary (Figure~7A, T4), where gravitational torque turns into a negative 
value (Figure~2f). Mutual scattering maintains separations of approximately $\Delta R$ (Figure~7A, T5). 
When the time scale of the orbital evolution, 
$\left( \tau_{r,\mathrm{planet}} = r_{\mathrm{planet}} / v_{r,\mathrm{planet}} \right)$, 
of all planets exceeds the time scale of disk evolution, 
$\bigl( \tau_{\mathrm{disk}} = 1 M_\odot / \dot{M} \bigr)$,
the calculations are stopped. In summary, for case A, three rocky planets are formed with the mass of 
$2.18 M_\oplus$, $2.26 M_\oplus$, and $0.46 M_\oplus$ (Table~1; Figure~7A, T4 and T5).

In case B (Table~1; Figure~7B), $M_{\mathrm{MRI}}$ increases from $2.49 \times 10^{-3}$ Myr to 
$M_{\mathrm{Rocky}} = 1.97 M_\oplus$ at $2.38 \times 10^{-2}$ Myr (Table~1). The first planet forms at 
$2.34 \times 10^{-3}$ Myr (Figure~7B, T1) and migrated out from the inner MRI front at $2.19 \times 10^{-2}$ 
Myr (Figure~7B, T2). Then, the first planet with the mass of $1.78 M_\oplus$, which is nearly equal to 
the analytically estimated mass ($1.6 M_\oplus$) by Equation~(\ref{eq:planetmass}). Second planet has mass $0.193 
M_\oplus$ from residual pebbles. In summary, for case B, two rocky planets are formed with the masses 
of $1.78 M_\oplus$ and $0.193 M_\oplus$ (Table~1; Figure~7B, T3 and T4).

In case C (Table~1; Figure~7C), $M_{\mathrm{MRI}}$ increases from $9.27 \times 10^{-4}$ Myr to 
$M_{\mathrm{Rocky}} = 1.96 M_\oplus$ at $9.39 \times 10^{-3}$ Myr (Table~1). A planet is formed at 
$8.22 \times 10^{-4}$ Myr (Figure~7C, T1) and grows until solid particles are completely used up, when 
its mass reaches $1.96 M_\oplus$. The planet migrated out from the inner MRI front at $2.19 
\times 10^{-2}$ Myr because its mass was small (Table~1; Figure~7C, T1 and T2).

In case D (Table~1; Figure~7D), $M_{\mathrm{MRI}}$ increases from $6.76 \times 10^{-3}$ Myr to 
$M_{\mathrm{Rocky}} = 1.99 M_\oplus$ at $6.17 \times 10^{-2}$ Myr (Figure~7D, T1). The first planet 
was formed at $6.62 \times 10^{-3}$ Myr (Figure~7D, T2). At $3.34 \times 10^{-2}$ Myr, the first planet 
migrated out from the inner MRI front, and the second planet was immediately formed (Figure~7D, T3). 
At $6.16 \times 10^{-2}$ Myr, the second planet migrates out from the inner MRI front, and the third 
planet is immediately formed (Figure~7D, T4). Then, the first and second planets with the mass of 
$0.974 M_\oplus$ and $1.04 M_\oplus$, which are nearly equal to the analytically estimated mass ($0.9 M_\oplus$) 
by Equation~(\ref{eq:planetmass}). A third planet with a mass of $1.15 \times 10^{-2} M_\oplus$ is formed from the residual pebbles. 
In summary, for case D, three rocky planets are formed with the mass of $0.974 M_\oplus$, $1.04 M_\oplus$, 
and $1.15 \times 10^{-2} M_\oplus$ (Table~1; Figure~7D, T5 and T6).

In case E (Table~1; Figure~7E), $M_{\mathrm{MRI}}$ increases from $6.12 \times 10^{-3}$ Myr to 
$M_{\mathrm{Rocky}} = 10.0 M_\oplus$ at $6.05 \times 10^{-2}$ Myr (Figure~7E, T1). 
The first planet was formed at $6.31 \times 10^{-3}$ Myr (Figure~7E, T2). 
At $6.71 \times 10^{-3}$ Myr, the first planet migrates out from the inner MRI front, and the second planet is immediately formed (Figure~7E, T2). 
At $2.02 \times 10^{-2}$ Myr, the second planet migrates out from the inner MRI front, and the third planet is immediately formed (Figure~7E, T3). 
At $5.09 \times 10^{-2}$ Myr, the third planet migrated out from the inner MRI front, and the fourth planet was immediately formed (Figure~7E, T4). 
In summary, for case E, four rocky planets are formed with the mass of $2.51 M_\oplus$, $2.64 M_\oplus$, 
$2.62 M_\oplus$, and $2.24 M_\oplus$ (Table~1; Figure~7E, T5 and T6), which are nearly equal to the analytically 
estimated mass ($2.3 M_\oplus$) by Equation~(\ref{eq:planetmass}). The simulation results (Table~2) are well explained by the analytical formula given in the Appendix.

\section{DISCUSSION} 
Terrestrial planet formation proceeds via three steps: 
1) Gravitational instability forms the first planet, which grows through pebble accretion. 
2) Sufficiently planetary massive triggers outward migration from the inner MRI front due to gravitational torque (Equations~(\ref{eq:gamma_0})--(\ref{eq:v_r_planet})). 
3) If the solid particles remain in the inner MRI front, a second planet forms immediately via gravitational instability. 
This cycle repeats until solid particles are exhausted.

When planet migrates from the inner MRI front, due to gas torque, their masses approximate Equation~(\ref{eq:planetmass}) values, comparable to Earth\textquotesingle s mass. The final planet, formed during pebble depletion, 
remains smaller than Earth-sized planet. All planets migrate from the inner MRI front and stabilize 
at the WSZ boundary where $\Gamma$ becomes negative (Figure~2f).

In case D, the resultant mass spectrum of the terrestrial planets was similar to that of the solar system; 
in other words, two Earth-mass planets (Earth and Venus) were successfully formed. 

In the remainder of this section, we discuss the chemical composition of the Earth, orbital elements 
of the planets, super-Earth formation, origin of Mars and Mercury, and Moon formation.

\subsection{Reductive Earth Compositions}
Terrestrial planets form at the inner MRI front, a reductive environment with temperature exceeding  $1000~\mathrm{K}$.
Consequently, Earth comprises reductive materials resembling enstatite chondrites. It is consistent with the following 
two facts. First, in the Urey-Craig diagram, Earth\textquotesingle s rocks are shown near the enstatite chondrites (e.g., \citealt{Yoshizaki2019}). 
Second, in studies of oxygen isotope ratios ($\delta^{17}$O and $\delta^{18}$O) 
\citep{Clayton1984, Javoy1995, Clayton1999, Javoy2010}, enstatite chondrites lie on the terrestrial fractionation line (TFL).
Because we can estimate the temperature, pressure, and dust/gas ratio where a planet was formed, 
we will calculate the chemical and mineral composition of the early Earth using the condensation theory 
\citep{Grossman1972, Yoneda1995, Ebel2000} as a future task.

Inner MRI front formation indicates terrestrial planets originated as dry bodies lacking atmospheric or oceanic components. 
Earth\textquotesingle s volatile inventory arrived later via bombardment of carbonaceous chondrites \citep{Maruyama2017}. The early Earth 
formed from reductive materials, creating a magma ocean that solidified under highly reducing, volatile-free conditions. 
During cooling, actinides (uranium and thorium), phosphorus (P), and potassium (K) concentrated in residual liquids due to 
their large ionic radii and incompatibility with mantle minerals. These residual melts produced KREEP (potassium (K), rare earth 
materials (REE), and phosphorous (P)) basalt, together with a buoyant anorthositic upper felsic crust. The primordial crust likely 
composed of anorthosite, KREEP basalt, and reductive iron minerals such as pyrite (FeS$_2$) and schreibersite (Fe$_3$P). 
These conditions suggest nuclear geyser as promising sites for life\textquotesingle s emergence, where underground natural reactors provided 
continuous material and energy flows \citep{Maruyama2017}.

\subsection{Orbital Element of Planets}
Case D, successfully reproduces solar system mass ratios (0.974 $M_{\oplus}$, 1.04 $M_{\oplus}$, and $1.15 \times 10^{-2} M_{\oplus}$) 
corresponding to Venus with 0.815 $M_{\oplus}$, Earth with 1.0 $M_{\oplus}$, and Mars with 0.107 $M_{\oplus}$. 
However, final orbital positions (1.54~AU, 1.39~AU, and 1.31~AU) differ significantly from observed values (Venus: 0.7233~AU) and (Earth: 1.0~AU). 
Long-term migration due to a decreasing accretion rates may explain this discrepancy.

\subsection{Possibility of Super-Earth and Hot Jupiter Formation}
Super-Earth possess masses several times Earth\textquotesingle s, while hot Jupiter are massive gas planets with a very short orbital period; 
in other words, the planet is located near the central star. Lower $\dot{M}$ increases migration timescale 
(Equation~(\ref{eq:migration_time_scale})), 
allowing planets to remain at the inner MRI front longer. Higher $f_{\mathrm{p}}$ increases surface density $\Sigma_{\mathrm{p}}$, accelerating 
planetary growth (see Equation~(\ref{eq:planetary_growth_rate})).

Observational surveys reveal multiple planet systems where some planets have nearly identical masses \citep{Lissauer2011}. 
This supports our model since planets forming in the same protoplanetary system experience similar conditions when migrating 
from the inner MRI front (Equations~(\ref{eq:migration_time_scale})–(\ref{eq:gamma_0})). Case E demonstrates multiple terrestrial planet formations representative of such systems.

When galactic cosmic ray ionization rates are high ($10^{-15} \ \mathrm{s}^{-1}$), the outer MRI front shifts inside the WSZ due to 
enhanced cosmic-ray flux \citep{Imaeda2017b}. The inner MRI front location remains unchanged as it depends primarily on thermal ionization. 
This configuration procedures super-Earths and hot Jupiters.

\subsection{Origin of Mars and Mercury}
Case D successfully forms two Earth-mass planets, but our solar system contains two additional, much smaller terrestrial planets: 
Mars and Mercury. We propose three hypotheses for their origins:

\subsubsection{Depletion Hypothesis}
The final planet remains small due to pebble depletion at the inner MRI front. This small ``depletion planet'' could represents 
Mars or Mercury\textquotesingle s origin, requiring substantial migration to current orbital positions.
\textbf{In particular, the radius (or diameter) ratio of Mars is strikingly similar to that of Earth and Venus
[Mars: $\sim 0.5414$ \citep{Archinal2018, Maistre2023}, Earth: $\sim 0.5462$ \citep{Archinal2018, Dziewonski1981},
Venus: $\sim 0.51--0.58$ \citep{Archinal2018, Amorim2023}].
Such similarity in radius ratios indicates that these bodies experienced comparable thermal conditions during their formation,
making it unlikely that they formed independently at distinct heliocentric locations, and instead suggesting formation at a similar heliocentric distance.
This observational constraint is fully consistent with the model, in which Earth and Venus form at the same location (the MRI front), while Mars forms from the remaining material, thereby strongly supporting this formation scenario.}

\subsubsection{Lagrange Point Hypothesis}
We \textbf{assumed} single planet formation per orbit, solid particles density via gravitational perturbation of this planet (Figure 7). 
However, two additional planets might form at Earth\textquotesingle s Trojan points (L4 and/or L5) which are gravitationally stable (Figure 10). 
As objects grow to approximately one-tenth Earth mass (Mars-sized), orbital instability initiates through gravitational interactions. 
One object impacts Earth (giant impact; see Section 4.5), while the other scatters from Earth\textquotesingle s orbit, potentially becoming Mars. 
Similar processes may have affected Venus.

\begin{figure}[htbp]
 \centering
 \includegraphics[width=0.8\linewidth]{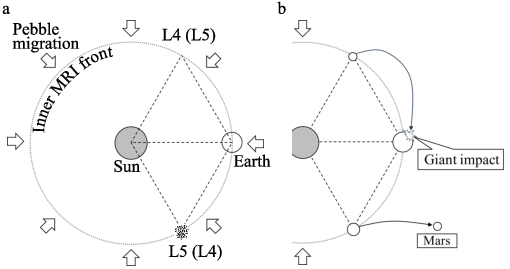}
 \caption{
Mars-sized object formation at Earth\textquotesingle s trojan points. First, the pebbles accumulated at Earth\textquotesingle s trojan points (a). 
Second, as the objects grow, the hill radius expand. When the five times of hill radius of the object are influence each other, 
relatively small object scatter out of system. On the other hand, relatively large object hits Earth (b).
 }
 \label{fig:figure10}
\end{figure}

\subsubsection{WSZ Hypothesis}
Mars-sized planets might form within the WSZ. Planet forming in the WSZ have smaller metallic cores than MRI front 
planets because the WSZ environment is more oxidizing. Martian rocks do not plot near enstatite chondrites in Urey-Craig diagrams 
\citep{Yoshizaki2020}, and SNC meteorites (Martian samples) lie above the terrestrial fractionation line in oxygen isotope space 
\citep{Clayton1984,Javoy1995,Clayton1999,Javoy2010}.

Beyond the WSZ, icy planetesimals form in the OTR through low relative velocity less than 1~cm s$^{-1}$ creating porous 
aggregations with low density down to $10^{-5}$~g~cm$^{-3}$ \citep{Okuzumi2012}. As aggregate mass increases, 
radial drift timescale becomes shorter than growth timescale. After pebbles drift into the QA and settle toward the midplane, 
forming an extremely thin subdisk, gravitational instability produces icy planetesimals at the outer MRI front.

\subsection{Moon Formation}
Giant impact theory proposes that a Mars-sized object collided with early Earth to form the Moon. This impactor may have 
formed simultaneously at the MRI front, as discussed in section 4.4.2. (Figure 10). \citet{Belbruno2005} suggested such a collision 
could produce lunar-mass satellites, consistent with giant impact scenarios.

\subsection{\textbf{Towards a Comprehensive Model of Solar System Formation}}
\textbf{In the present model, only a single planet forms at each orbital location, and the potential formation of additional bodies 
at the Lagrange points (L4 and/or L5) has not been considered. Co-orbital planets could grow at these points, 
potentially influencing the evolution of planets formed elsewhere in the system. 
Furthermore, possible orbital rearrangements or migrations among the formed planets were not taken into account, 
even though such processes could modify the final orbital architecture and mass distribution.}

\textbf{Future research should address the formation and migration of giant planets in conjunction with the dissipation of the gaseous disk. 
Additionally, the fragmentation velocity of solid particles near the water sublimation zone (WSZ) must be considered, 
as this region may affect the formation conditions of rocky planets and asteroid belts. 
The current tandem planet formation scenario has also yet to include the formation of the Edgeworth–Kuiper belt and the Oort cloud.}

\textbf{Incorporating these processes is essential for developing a comprehensive, integrated model of solar system formation.}

\section{CONCLUSIONS} 
We present a terrestrial planet formation theory demonstrating that Earth-mass planets form naturally in tandem protosolar disks 
based on tandem planet formation theory \citep{Ebisuzaki2017, Imaeda2017a, Imaeda2017b, Imaeda2018} In our model, planets 
form at the inner MRI front where pebbles accumulate between this boundary and the inner WSZ edge. Planet inevitably 
migrates when the reaching Earth-size mass due to gravitational torque.

In the future, we will estimate the chemical and mineral composition of the early Earth using the condensation theory 
(\citealp{Grossman1972, Yoneda1995, Ebel2000}) as a future task. It is important to consider the birthplace of primitive life.

The future task will be orbital formation after the planets are formed. This scheme can solve exoplanet formation theories 
such as super-Earth and hot-Jupiter.

\appendix
\renewcommand{\thefigure}{A\arabic{figure}}
\setcounter{figure}{0}
The Appendix provides an estimate of the planet mass, as described in Equation~(\ref{eq:planetmass}) in the main text.
\vspace{2ex}
\section{APPROXIMATE MIDPLANE TEMPERATURE IN THE TERRESTRIAL PLANET FORMATION REGION}
An irradiation temperature, $T_{\mathrm{irr}}$, is given by:
\begin{equation}
T_{\mathrm{irr}}^4 = \frac{1}{2} (1-\epsilon) T_*^4 \left(\frac{R_*}{r}\right)^2 \left[ \frac{4}{3\pi} \left(\frac{R_*}{r}\right) + \frac{2}{7} \frac{H}{r} \right],
\end{equation}
where $\epsilon = 0.5$ is the disk albedo \citep{Coleman2014}, $T_* = 4000\,\mathrm{K}$ is the stellar temperature, 
and $R_* = 3 R_{\odot}$ is the stellar radius. Here, $r$ and $H$ are the semi-major axis and vertical-scale height of 
the disk, respectively. In the rocky planet formation region, because $ r \gg R_* $, the stellar irradiation temperature can 
be neglected ($T_{\mathrm{irr}} \simeq 0$). The optical depth $\tau$ is assumed to be greater than 500 
(Eq.~26 in \citealt{Ebisuzaki2017}; see also \citealt{Imaeda2017a}; hereafter EI). The midplane temperature, $T_{\mathrm{m}}$,
 can then be approximated 
using Equation~(8) from EI as follows:
\begin{eqnarray}
T_{\mathrm{m}}^4 &\simeq& \left( \frac{3}{2^3} \right)^2 \frac{\dot{M} \, \Omega^2 \, \tau}{\pi \, \sigma},
\label{eq:Tm4a}
\end{eqnarray}
where $\Omega$ and $\sigma$ are the Keplerian orbital frequency and the Stefan–Boltzmann constant, respectively. 
$\dot{M}$ is the accretion rate of the protosolar disk. The optical depth, $\tau$, which appears in Equation~(\ref{eq:Tm4a}), is given by:
\begin{eqnarray}
\tau = \frac{\kappa \gamma \Sigma}{2}
\label{eq:optical_depth},
\end{eqnarray}
(EI Eq. 26), where $\gamma = 10^{-0.5}$ (EI). Here, $\kappa$ and $\Sigma$ are the opacity and the column density 
for MRI inactive region, respectively, which are defined as:
\begin{eqnarray}
\kappa &=& \beta_{\kappa} T_{\mathrm{m}}^{3/4}, \label{eq:kappa}\\
\Sigma &=& \frac{\dot{M} \Omega}{3 \pi \bar{\alpha} c_{ \mathrm{s} }^2}, \label{eq:Sigma}
\end{eqnarray}
where $\beta_{\kappa} = 2.13 \times 10^{-2}\ \mathrm{cm}^2\,\mathrm{g}^{-1}\,\mathrm{K}^{-3/4}$ \citep{Stepinski1998}. 
$c_{\mathrm{s}}$ is the isothermal sound velocity. $\bar{\alpha}$ is the $\alpha$-value in the $\alpha$-model \citep{Shakura1973}. 
In the MRI suppressed region, $\bar{\alpha}$ is defined as $\gamma \alpha_{\mathrm{act}}$, where $\alpha_{\mathrm{act}} = 1.0 \times 10^{-2}$
represents the $\alpha$-value in the MRI active region \citep{Davis2010, Shi2010}. $\gamma = 10^{-0.5}$ 
is the reduction factor of turbulence due to the suppression of MRI (EI). From Equations~(\ref{eq:Tm4a})–(\ref{eq:Sigma}):
\begin{equation}
T_{\mathrm{m}}^4 \simeq \frac{3}{2^7 \pi^2} \frac{\dot{M}^2 \Omega^3 \beta_\kappa T_{\mathrm{m}}^{3/4}}{\alpha_{\mathrm{act}} \sigma c_{\mathrm{s}}^2}.
\label{eq:Tm4b}
\end{equation}
From Equations~(\ref{eq:Tm4b}), and Equations~(6), and (7) in EI, the approximate midplane temperature in the rocky planet formation 
region can be estimated as:
\begin{eqnarray}
T_{\mathrm{m}} \simeq \left[ \frac{3}{2^7 \pi^2} \frac{\dot{M}^2 \beta_{\kappa} (\mu m_{\mathrm{H}}) (G M_*)^{3/2}}{\alpha_{\mathrm{act}} \sigma k_{\mathrm{B}} r^{9/2}} \right]^{4/17},
\label{eq:midplane_temperature}
\end{eqnarray}
where $k_{\mathrm{B}}$ is the Boltzmann constant, $\mu=2.34$ is the mean molecular weight of the gas, $m_{\mathrm{H}}$ is the mass of a 
hydrogen atom, $G$ is the gravitational constant, and $M_*$ is the mass of the central star.

\section{AN APPROXIMATE ESTIMATION OF $\eta$ IN THE TERRESTRIAL PLANET FORMATION REGION}
\renewcommand{\theequation}{B\arabic{equation}}  
\setcounter{equation}{0}                       

The gas density at the midplane $\rho_{\mathrm{m}}$ is given by:
\begin{eqnarray}
\rho_{\mathrm{m}} = \frac{\Sigma}{\sqrt{2\pi} H}. \label{eq:midplane_density_a}
\end{eqnarray}
By combining Equations~(\ref{eq:Sigma}), (\ref{eq:midplane_density_a}), and Equation~(5) in EI:
\begin{eqnarray}
\rho_{\mathrm{m}} &=& \frac{1}{2^{1/2} \times 3 \times \pi^{3/2}} \frac{\dot{M} \Omega^2}{\gamma \alpha_{\mathrm{act}} c_{\mathrm{s}}^3}. \label{eq:midplane_density_b}
\end{eqnarray}
Utilizing Equation~(\ref{eq:midplane_density_b}) along with Equations~(6) and (7) in EI:
\begin{eqnarray}
\rho_{\mathrm{m}} T_{\mathrm{m}} = \frac{1}{2^{1/2} \times 3 \times \pi^{3/2}} \frac{\dot{M}}{\gamma \alpha_{\mathrm{act}}} \left(\frac{\mu m_{\mathrm{H}}}{k_{\mathrm{B}}}\right)^{3/2} \frac{G M_*}{r^3} T_{\mathrm{m}}^{-1/2}. \label{eq:rhom_Tm_a}
\end{eqnarray}
Further, substituting Equations~(\ref{eq:midplane_temperature}) and (\ref{eq:rhom_Tm_a}):
\begin{eqnarray}
\rho_{\mathrm{m}} T_{\mathrm{m}} \simeq \left[
\left(\frac{2^{11}}{3^{38} \pi^{43}}\right)^{1/2} 
\frac{\dot{M}^{13} \sigma^2}{\gamma^{17} \alpha_{\mathrm{act}}^{15} \beta_{\kappa}^2} 
\left(\frac{\mu m_{\mathrm{H}}}{k_{\mathrm{B}}}\right)^{47/2} (G M_*)^{14}
\right]^{1/17} r^{-42/17}. \label{eq:rhom_Tm_b}
\end{eqnarray}
Therefore, according to Equation~(\ref{eq:rhom_Tm_b}):
\begin{eqnarray}
\frac{\partial \log_{10}(\rho_{\mathrm{m}} T_{\mathrm{m}})}{\partial \log_{10} r} \simeq -\frac{42}{17}. \label{eq:rhoT_radial_grad}
\end{eqnarray}
As a result, the parameter $\eta$, as defined by EI, is expressed from Equation~(\ref{eq:rhoT_radial_grad}) as follows: 
\begin{equation}
\eta \simeq - \frac{c_{\mathrm{s}}^2}{2 r^2 \Omega^2} \frac{\partial \log_{10} (\rho_{\mathrm{m}} T_{\mathrm{m}})}{\partial \log_{10} r} 
= \frac{21}{17} \frac{c_{\mathrm{s}}^2}{r^2 \Omega^2}.
\label{eq:eta}
\end{equation}

\section{PARTICLE STOPPING TIME IN THE EPSTEIN DRAG REGIME}
\setcounter{equation}{0}
\renewcommand{\theequation}{C\arabic{equation}}
The thermal velocity of a superparticle, $v_{\mathrm{th},k}$, which is defined as:
\begin{eqnarray}
v_{\mathrm{th},k} &=& \left( \frac{2^3}{\pi} \right)^{1/2} c_{\mathrm{s},k}, \label{eq:vth}
\end{eqnarray}
where $k$ is the superparticle index. $c_{\mathrm{s},k}$ is the isothermal sound speed at the location of superparticle $k$. 
From Equation~(\ref{eq:vth}) and Equations~(64) and (65) in EI, the stopping time of a particle in the case of the Epstein Drag Regime, $t_{\mathrm{s},k}$, is given by:
\begin{eqnarray}
t_{\mathrm{s},k} &=& \left(\frac{3^2}{2^7 \pi}\right)^{1/2} \frac{m_{\mathrm{p},k}}{a_{p,k}^2 \rho_k c_{\mathrm{s},k}}. \label{eq:stoping_time}
\end{eqnarray}
Here, $m_{\mathrm{p},k}$ and $a_{\mathrm{p},k}$ are the average mass and radius of superparticle $k$, respectively. 
$\rho_k$ is the gas density at the location of superparticle $k$.

\section{Velocity of Mass Accretion to the Inner MRI Front (\protect\lowercase{$v_{\mathrm{mp}}$})}

\setcounter{equation}{0}
\renewcommand{\theequation}{D\arabic{equation}}

Figure A1 shows the time evolution of various velocity components for the case of $k = 4$: particle-particle relative velocity, 
$v_{\mathrm{rel,pp},k}$, Brownian motion, $v_{\mathrm{B},k}$, radial drift difference, $v_{r\mathrm{pp},k}$, and vertical settling 
difference velocity, $v_{z\mathrm{pp},k}$, as defined by Equations~(66)–(68) and (70) in EI. 
When $\Omega t_{\mathrm{s}} \ll 1$, the radial drift difference, $v_{r\mathrm{pp},k}$, azimuthal drift difference 
$v_{\mathrm{\phi pp},k}$, (Equations~(68) and (69) in EI) can be approximated as follows:
\begin{eqnarray}
v_{r\mathrm{pp},k} &\simeq& \Omega_k^2 t_{\mathrm{s},k} \eta_k r_{\mathrm{p},k}, \label{eq:vrpp} \\
v_{\mathrm{\phi pp},k} &\simeq& 0. \label{eq:vphipp}
\end{eqnarray}
Here, $\Omega_k$, $\eta_k$ and $r_{\mathrm{p},k}$ are the Keplerian orbital frequency, the parameter $\eta$, and the distance 
from the central star at the location of superparticle $k$, respectively.

\begin{figure}[htbp]
 \centering
 \includegraphics[width=0.8\linewidth]{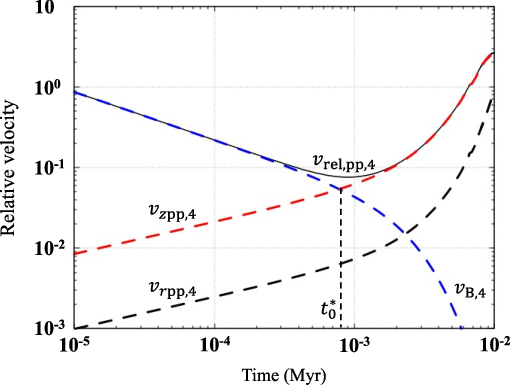}
 \caption{The evolution of the particle-particle relative velocity $v_{\mathrm{rel,pp},k}$ (solid line), Brownian motion $v_{\mathrm{B},k}$ (blue dashed line), 
radial drift difference $v_{r\mathrm{pp},k}$ (black dashed line), and vertical settling difference velocity $v_{z\mathrm{pp},k}$ (red dashed line), 
for accretion rate of $\dot{M} = 10^{-7.08} M_\odot\,\mathrm{yr}^{-1}$.
as defined by Equations~(66)--(68) and (70) in EI, for the case of $k=4$. The vertical dashed lines indicate $t_{0,k}^*$ (Equation~(\ref{eq:time_vzpp_equals_vB_b})), 
the time when $v_{z\mathrm{pp},k} = v_{\mathrm{B},k}$.
}
 \label{fig:figureA01}
\end{figure}

From $\Omega_k t_{\mathrm{s},k} \ll 1$ and Equation~(48) in EI, the vertical settling velocity, $v_{z\mathrm{p},k}$, is approximated as:
\begin{eqnarray}
v_{z\mathrm{p},k} &\simeq& 0. \label{eq:vzp}
\end{eqnarray}
Equation~(\ref{eq:vzp}) shows that the scale height of the particles, $z_{\mathrm{p},k}$, initially retains the vertical scale height of the 
disk at the location of the superparticle $k$, denoted as $H_k$, according to the following relation:
\begin{eqnarray}
z_{\mathrm{p},k} &\simeq& H_k. \label{eq:zp}
\end{eqnarray}
From Equation~(\ref{eq:zp}) and Equation~(70) in EI, the vertical settling drift velocity, $v_{z\mathrm{pp},k}$, can be approximated as follows:
\begin{eqnarray}
v_{z\mathrm{pp},k} &\simeq& \frac{1}{2} \Omega_k^2 t_{\mathrm{s},k} H_k. \label{eq:vzpp}
\end{eqnarray}
In the QA, when the particle density at the midplane, $\rho_{\mathrm{pm},k}$, is lower than the gas density, $\rho_{\mathrm{m},k}$, the turbulent velocity, 
$v_{\mathrm{turb,pp},k}$, (Equation~(71) in EI) vanishes. Consequently, the viscous parameter, $\alpha_{\mathrm{D},k}$, (Equation~(71) in EI) is zero. 
In the absence of massive bodies in orbit, gravitational perturbations to solid particles are negligible. Thus, the viscous stirring velocity, $v_{\mathrm{VS}}$, (Equation~(75) in EI) is much 
 smaller than 1. From Equations~(\ref{eq:eta}), (\ref{eq:vrpp}), (\ref{eq:vzpp}), and Equation~(5) in EI,
The ratio $v_{r\mathrm{pp},k} / v_{z\mathrm{pp},k}$ is given by:
\begin{eqnarray}
\frac{v_{r\mathrm{pp},k}}{v_{z\mathrm{pp},k}} &\simeq& \frac{2 \times 21}{17} \frac{H_k}{r_{\mathrm{p},k}}.
\end{eqnarray}
Since $H_k / r_{\mathrm{p},k} \ll 1$, it follows that:
\begin{eqnarray}
\frac{v_{z\mathrm{pp},k}}{v_{r\mathrm{pp},k}} &\gg& 1. \label{eq:vz_over_vr}
\end{eqnarray}
When $v_{z\mathrm{pp},k} = v_{B,k}$, the initial mass of a superparticle, $m_{0,k}^*$ is defined, and it follows from Equations~(\ref{eq:stoping_time}) and (\ref{eq:vzpp}), 
and Equations~(5), (50) and (67) in EI that:
\begin{eqnarray}
m_{0,k}^* &\simeq& \left( 
\frac{
2^{31} \rho_k^{6} \left(k_{\mathrm{B}} T_{\mathrm{m},k}\right)^3
}{
3^{2} \pi^{4} \rho_{\mathrm{i},k}^{4} \Omega_k^{6}
} 
\right)^{1/5}, \label{eq:initial_mass_superparticle_a}
\end{eqnarray}
where $\rho_{\mathrm{i},k}$ is the internal density of superparticle  $k$. In the terrestrial planet formation region, $\rho_{\mathrm{i},k}$ is assumed to be $2\,\mathrm{g\,cm^{-3}}$, 
representing rocky density. Here:
\begin{eqnarray}
\rho_k = e^{-1/2} \rho_{\mathrm{m},k}. \label{eq:gas_density}
\end{eqnarray}
From Equations~(\ref{eq:midplane_temperature}), (\ref{eq:midplane_density_b}), (\ref{eq:initial_mass_superparticle_a}), (\ref{eq:gas_density}), and Equation~(5) and (6) for EI:
\begin{eqnarray}
m_{0,k}^* &\simeq& \left[
\frac{2^{42} e^{-3}}{3^{10} \pi^{9}} 
\frac{
\dot{M}^2 \sigma^2 
T_{\mathrm{m}}^{5/2} \left( \mu m_{\mathrm{H}} \right)^7
}{
\rho_{\mathrm{i},k}^4 \gamma^6 \alpha_{\mathrm{act}}^4 \beta_{\kappa}^2 k_{\mathrm{B}}^4
}
\right]^{1/5}.
\label{eq:initial_mass_superparticle_b}
\end{eqnarray}
From Equations~(\ref{eq:stoping_time}) and (\ref{eq:vzpp}), and Equations~(5), (50), and (67) in EI:
\begin{eqnarray}
\frac{v_{B,k}}{v_{z\mathrm{pp},k}} &\simeq& \left(\frac{2^{31}}{3^{2} \pi^{4}}\right)^{1/6} 
\frac{\rho_k (k_{\mathrm{B}} T_{\mathrm{m},k})^{1/2}}
{\rho_{\mathrm{i},k}^{2/3} \Omega_k} m_{\mathrm{p},k}^{-5/6}. \label{eq:vB_over_vzpp}
\end{eqnarray}
Thus, at $t = t_{0,k}^*$, the mass is $m_{0,k}^*$. When $t \leq t_{0,k}^*$, then $m_{p,k} \leq m_{0,k}^*$. Therefore:
\begin{eqnarray}
v_{B,k} &\geq& v_{z\mathrm{pp},k}. \label{eq:vB_geq_vzpp}
\end{eqnarray}

From Equations~(\ref{eq:vphipp}), (\ref{eq:vz_over_vr}), and (\ref{eq:vB_geq_vzpp}), together with $v_{\mathrm{turb,pp},k} = 0$ and $v_{\mathrm{VS}} \ll 1$, 
the relative velocity, $v_{\mathrm{rel,pp},k}$, is approximately equal to $v_{B}$ by Equation~(66) in EI, when $t \leq t_{0,k}^*$. Here, from Equation~(\ref{eq:initial_mass_superparticle_b}) and Equations~(49), (50), and (67) in EI:
\begin{eqnarray}
\left(\frac{v_{\mathrm{esc},k}}{v_{\mathrm{B},k}}\right)^2 &\leq& 
\left[
\frac{2^{35} e^{-3}}{3^{11} \pi^{5}}
\frac{
\dot{M}^2 \sigma^2 (\mu m_{\mathrm{H}})^7 G^3
}{
\rho_{\mathrm{i},k}^3 \gamma^6 \alpha_{\mathrm{act}}^4 \beta_{\kappa}^2 k_B^7 T_{\mathrm{m},k}^{1/2}
}
\right]^{1/3} \nonumber \\
&\simeq& 
(2.20 \times 10^{-11}) \left(\frac{\dot{M}}{10^{-7.08} M_{\odot} \mathrm{yr}^{-1}}\right)^{2/3} 
\left(\frac{T_{\mathrm{m},k}}{1000\, \mathrm{K}}\right)^{-1/6}, \label{eq:ineq_vesc_vB_squared_a}
\end{eqnarray}
where $v_{\mathrm{esc},k}$ are the escape velocities of the particles. Therefore, if $m_{\mathrm{p},k} \leq m_{0,k}^*$, 
then $(v_{\mathrm{esc},k} / v_{\mathrm{B},k})^2 \ll 1.$ From Equation (46) in EI, and $v_{\mathrm{rel,pp},k} \simeq v_{\mathrm{B}}$:
\begin{equation}
\frac{\partial m_{\mathrm{p},k}}{\partial t} \simeq \pi a_{\mathrm{p},k}^2 \rho_{\mathrm{p},k} v_{\mathrm{B},k}.
\label{eq:mass_growth_rate_vB}
\end{equation}
Here, $\rho_{\mathrm{p},k}$ is the particle density at the particle scale height, and using $f_p$ (particle fraction) it becomes:
\begin{eqnarray}
\rho_{\mathrm{p},k} = f_{\mathrm{p}} \rho_k. \label{eq:particle_density}
\end{eqnarray}
We use Equation~(\ref{eq:particle_density}) in this work,  
together with Equations~(50) and (67) in EI.  
By integrating Equation~(\ref{eq:mass_growth_rate_vB}) over $(t)$ from $t=0$ to $t=t_{0,k}^*$ 
and over $m_{\mathrm{p},k}$ from $m_{\mathrm{p},k}=0$ to $m_{\mathrm{p},k}=m_{0,k}^*$, we obtain:
\begin{equation}
t_{0,k}^* \simeq \frac{(2 \times 3 \times \pi^{1/2})^{1/3}}{5} 
\frac{\rho_{\mathrm{i},k}^{2/3} \, m_{0,k}^{*5/6}}{f_{\mathrm{p}} \, \rho_k \, (k_{\mathrm{B},k} \, T_{\mathrm{m},k})^{1/2}}.
\label{eq:time_vzpp_equals_vB_a}
\end{equation}
From Equations (\ref{eq:initial_mass_superparticle_a}) and (\ref{eq:time_vzpp_equals_vB_a}):
\begin{equation}
t_{0,k}^* \simeq \frac{2^{11/2}}{5 \pi^{1/2} f_{\mathrm{p}} \, \Omega_k}.
\label{eq:time_vzpp_equals_vB_b}
\end{equation}
Thus, if $t \geq t_{0,k}^*$, then $m_{p,k} \geq m_{0,k}^*$. Therefore, form Equation (\ref{eq:vB_over_vzpp}):
\begin{equation}
v_{z\mathrm{pp},k} \geq v_{\mathrm{B},k}.
\label{eq:vzpp_geq_vB}
\end{equation}
From Equations (\ref{eq:vphipp}), (\ref{eq:vz_over_vr}), and (\ref{eq:vzpp_geq_vB}), together with $v_{\mathrm{turb,pp},k} = 0$ and $v_{\mathrm{VS}} \ll 1$, 
the $v_{\mathrm{rel,pp},k}$ is approximately $v_{\mathrm{zpp},k}$ by Equation (66) in EI, when $t \geq t_0^*$. 
Here, from Equation (\ref{eq:midplane_temperature}), (\ref{eq:midplane_density_b}), (\ref{eq:stoping_time}), (\ref{eq:vzpp}), (\ref{eq:gas_density}), and Equations (5), (6), (7), (49), and (50) in EI:
\begin{eqnarray}
\left(\frac{v_{\mathrm{esc},k}}{v_{z\mathrm{pp},k}}\right)^2 &\leq& 
\left[
\frac{2^{35} e^{-3}}{3^{11} \pi^{5}}
\frac{
\dot{M}^2 \sigma^2 (\mu m_{\mathrm{H}})^7 G^3
}{
\rho_{\mathrm{i},k}^3 \gamma^6 \alpha_{\mathrm{act}}^4 \beta_{\kappa}^2 k_B^7 T_{\mathrm{m},k}^{1/2}
}
\right]^{1/3} \nonumber \\
&\simeq& 
(2.20 \times 10^{-11}) \left(\frac{\dot{M}}{10^{-7.08} M_{\odot} \mathrm{yr}^{-1}}\right)^{2/3} 
\left(\frac{T_{\mathrm{m},k}}{1000\, \mathrm{K}}\right)^{-1/6}.\label{eq:ineq_vesc_vB_squared_b}
\end{eqnarray}
The right sides of Equations~(\ref{eq:ineq_vesc_vB_squared_a}) and (\ref{eq:ineq_vesc_vB_squared_b}) are identical. Therefore:
\begin{eqnarray}
\left( \frac{v_{\mathrm{esc},k}}{v_{z\mathrm{pp},k}} \right)^2 &\ll& 1. \label{eq:ineq_vesc_vzpp_squared_ll1}
\end{eqnarray}

From Equation~(\ref{eq:ineq_vesc_vzpp_squared_ll1}), and Equation~(46) in EI, and $v_{\mathrm{rel,pp},k} \simeq v_{z\mathrm{pp},k}$:
\begin{eqnarray}
\frac{\partial m_{\mathrm{p},k}}{\partial t} &\simeq& \pi a_{\mathrm{p},k}^2 \rho_{\mathrm{p},k} v_{z\mathrm{pp},k}. \label{eq:mass_growth_rate_vzpp}
\end{eqnarray}
From Equation~(\ref{eq:stoping_time}), (\ref{eq:vzpp}), (\ref{eq:particle_density}), (\ref{eq:mass_growth_rate_vzpp}) and Equation~(5) for EI:
\begin{eqnarray}
\frac{\partial m_{\mathrm{p},k}}{\partial t} &\simeq& \left( \frac{3^2 \pi}{2^9} \right)^{1/2} f_{\mathrm{p}} \, \Omega_k \, m_{\mathrm{p},k}. \label{eq:mass_growth_rate}
\end{eqnarray}
By integrating Equation~(\ref{eq:mass_growth_rate}) over $(t)$ from $t_{0,k}^*$ to $t_k^*$ and $m_{\mathrm{p},k}$ from $m_{0,k}^*$ to $m_{\mathrm{p},k}^*$, where $t_k^*$ and $m_{\mathrm{p},k}^*$ 
are the time and mass values at the end of the integration, respectively.
\begin{eqnarray}
\ln\left(\frac{m_{\mathrm{p},k}^*}{m_{0,k}^*}\right)
&\simeq& \left(\frac{3^2 \pi}{2^9}\right)^{1/2} f_{\mathrm{p}} \Omega_k (t_k^* - t_{0,k}^*).
\label{eq:ln_mass_growth}
\end{eqnarray}
From Equations~(\ref{eq:time_vzpp_equals_vB_b}), (\ref{eq:ln_mass_growth}), and Equation~(7) in EI:
\begin{eqnarray}
r_{\mathrm{p},k} 
&\simeq& \left(\frac{3^2 \pi}{2^9}\right)^{1/3} 
\left\{ \ln\left(\frac{m_{\mathrm{p},k}^*}{m_{0,k}^*}\right) + \frac{2 \times 3}{5} \right\}^{-2/3} 
\left(f_{\mathrm{p}} t_k^*\right)^{2/3} (G M_*)^{1/3}.
\label{eq:approx_rp_logmass_time}
\end{eqnarray}
From Equation~(\ref{eq:approx_rp_logmass_time}), the solid particles at $r_{\mathrm{p},k}$ accrete to the MRI front with a velocity $v_{\mathrm{mp}}$:
\begin{eqnarray}
v_{\mathrm{mp}} &\simeq& \frac{\partial r_{\mathrm{p},k}}{\partial t_k^*} = \frac{2}{3} r_{\mathrm{p},k} t_k^{*-1}. \label{eq:vmp}
\end{eqnarray}

\section{Mass of the Particle after $H_{\lowercase{\mathrm{k}}}/2$ Migration (\protect\lowercase{$m_{\mathrm{p}}^*$})}
\setcounter{equation}{0}
\renewcommand{\theequation}{E\arabic{equation}}

Because $ \rho_{\mathrm{p m}, \mathrm{k}} < \rho_{\mathrm{m}, \mathrm{k}} $, the radial drift velocity due to the turbulence by Kelvin-Helmholtz instability in the subdisk, 
$v_{r\mathrm{KH},k}$, is zero. Therefore, the radial drift velocity, ${\partial r_{\mathrm{p},k}}/{\partial t}$, is given by Equation~(47) in EI as follows:
\begin{eqnarray}
\frac{\partial r_{\mathrm{p},k}}{\partial t} &\simeq& 2 \Omega_k^2 t_{\mathrm{s},k} \eta r_{\mathrm{p},k}. \label{eq:radial_drift_velocity_a}
\end{eqnarray}
From Equations~(\ref{eq:eta}) and (\ref{eq:radial_drift_velocity_a}):
\begin{eqnarray}
\frac{\partial r_{\mathrm{p},k}}{\partial t} &\simeq& \frac{2 \times 3 \times 7}{17} 
\frac{c_{\mathrm{s},k}^2 t_{\mathrm{s},k}}{r_{\mathrm{p},k}}. \label{eq:radial_drift_velocity_b}
\end{eqnarray}
Using Equations~(\ref{eq:vzpp}), (\ref{eq:mass_growth_rate_vzpp}), (\ref{eq:radial_drift_velocity_b}), Equation~(5) in EI:
\begin{eqnarray}
\frac{\partial m_{\mathrm{p},k}}{\partial r_{\mathrm{p},k}} 
&\simeq& \frac{17 \pi}{2^{2} \times 3 \times 7} 
a_{\mathrm{p},k}^2 \, \rho_{\mathrm{p},k} \left(\frac{r_{\mathrm{p},k}}{H_k}\right). \label{eq:approx_dm_dr_a}
\end{eqnarray}
From Equations~(\ref{eq:Sigma}), (\ref{eq:midplane_density_a}), (\ref{eq:gas_density}), (\ref{eq:particle_density}), (\ref{eq:approx_dm_dr_a}), and Equations~(5) and (50) in EI:
\begin{eqnarray}
\frac{\partial m_{\mathrm{p},k}}{\partial r_{\mathrm{p},k}} 
&\simeq& \frac{17 e^{-1/2}}{2^{23/6} \times 3^{4/3} \times 7 \pi^{7/6}} 
\frac{f_{\mathrm{p}} \dot{M} r_{\mathrm{p},k}}{\gamma \alpha_{\mathrm{act}} H_k^{3} c_{\mathrm{s},k}} 
\left(\frac{m_{\mathrm{p},k}}{\rho_{\mathrm{i},k}}\right)^{2/3}. \label{eq:approx_dm_dr_b}
\end{eqnarray}
By integrating Equation~(\ref{eq:approx_dm_dr_b}) over $r_{\mathrm{p},k}$ from $r_{\mathrm{p},k,0}$ (the initial value of $r_{\mathrm{p},k}$) to $r_{\mathrm{p},k,0} + H_k/2$
 and $m_{\mathrm{p},k}$ from 0 to $m_{\mathrm{p},k}^*$:
\begin{eqnarray}
m_{\mathrm{p},k}^* &\simeq&
\left[
\frac{
17 e^{-1/2}
}{
2^{29/6} \times 3^{7/3} \times 7 \pi^{7/6}
}
\frac{
f_{\mathrm{p}} \dot{M}
}{
\rho_{\mathrm{i},k}^{2/3} \, \gamma \alpha_{\mathrm{act}} \, H_k^3 \, c_{\mathrm{s},k}
}
\left\{
\left( r_{\mathrm{p},k,0} + \frac{H_k}{2} \right)^2 - r_{\mathrm{p},k,0}^2
\right\}
\right]^3.
\label{eq:mp_star_from_radial_integ_a}
\end{eqnarray}
Here:
\begin{eqnarray}
\left( r_{\mathrm{p},k,0} + \frac{H_k}{2} \right)^2 - r_{\mathrm{p},k,0}^2
&\simeq&
H_k r_{\mathrm{p},k,0}. \label{eq:rk_square_diff_approx}
\end{eqnarray}
From Equations~(\ref{eq:midplane_temperature}), (\ref{eq:mp_star_from_radial_integ_a}), (\ref{eq:rk_square_diff_approx}), and Equations~(5) and (6) for EI:
\begin{eqnarray}
m_{\mathrm{p},k}^* &\simeq& 
\left[
\frac{
17^{18} e^{-9}
}{
2^{31} \times 3^{50} \times 7^{18} \pi^{5}
}
\left\{
\frac{
f_{\mathrm{p}}^{18} \dot{M}^2 \sigma^{8} T_{\mathrm{m}}^{7} (\mu m_{\mathrm{H}})^{19} (G M_*)^{6}
}{
\rho_{\mathrm{i},k}^{12} \gamma^{18} \alpha_{\mathrm{act}}^{10} \beta_{\kappa}^{8} k_{\mathrm{B}}^{19}
}
\right\}
\right]^{1/6}.
\label{eq:mp_star_from_radial_integ_b}
\end{eqnarray}
The ratio of Equations~(\ref{eq:initial_mass_superparticle_b}) and (\ref{eq:mp_star_from_radial_integ_b}) gives:
\begin{eqnarray}
\frac{m_{\mathrm{p},k}^*}{m_{0,k}^*} &=&
\left[
\frac{
17^{90} \times \pi^{29} \times \mathrm{e}^{-27}
}{
2^{407} \times 3^{190} \times 7^{90}
}
\frac{
f_{\mathrm{p}}^{90} \, \sigma^{28} \, T_{\mathrm{m}}^{20} \, (\mu m_{\mathrm{H}})^{53} \, (G M_*)^{30}
}{
\dot{M}^2 \, \gamma^{54} \, \alpha_{\mathrm{act}}^{26} \, \rho_{\mathrm{i},k}^{36} \, \beta_k^{28} \, k_{\mathrm{B}}^{71}
}
\right]^{1/30}.
\end{eqnarray}
Therefore:
\begin{eqnarray}
\log_{10} \left( \frac{m_{\mathrm{p},k}^*}{m_{0,k}^*} \right) &=&
\frac{1}{30} \log_{10} \left[
\frac{
(1.25 \times 10^{-3})^{90} \times 17^{90} \times \pi^{29} \times e^{-27}
}{
2^{407} \times 3^{190} \times 7^{90}
}
\frac{
\sigma^{28} \, (1000\,\mathrm{K})^{20} \, (\mu m_{\mathrm{H}})^{53} \, (G M_\odot)^{30}
}{
(10^{-7.08} \, M_\odot\,\mathrm{yr}^{-1})^2 \, \gamma^{54} \, \alpha_{\mathrm{act}}^{26} \, \rho_{{\mathrm{i}},k}^{36} \, \beta_k^{28} \, k_{\mathrm{B}}^{71}
}
\right] 
\nonumber \\[1ex]
&&
+ 3 \log_{10} \left( \frac{f_{\mathrm{p}}}{1.25 \times 10^{-3}} \right) 
+ \log_{10} \left( \frac{M_*}{M_\odot} \right) 
- \frac{1}{15} \log_{10} \left( \frac{\dot{M}}{10^{-7.08} \, M_\odot\,\mathrm{yr}^{-1}} \right) 
+ \frac{2}{3} \log_{10} \left( \frac{T_{\mathrm{m}}}{1000\,\mathrm{K}} \right) 
\nonumber \\[1ex]
&\simeq&
8.4 + 3 \log_{10} \left( \frac{f_{\mathrm{p}}}{1.25 \times 10^{-3}} \right).
\label{eq:log_mass_ratio_scaling}
\end{eqnarray}

\section{RELATIONSHIP BETWEEN MASS ACCRETION RATE TO THE INNER MRI FRONT AND PLANETARY MASS}
\setcounter{equation}{0}
\renewcommand{\theequation}{F\arabic{equation}}

The mass accretion rate to the inner MRI front can be approximated by:
\begin{eqnarray}
\frac{dM_{\mathrm{planet}}}{dt} &\simeq& \dot{M}_{\mathrm{MRI,total}},
\label{eq:planet_accretion_rate}
\end{eqnarray}
where $M_{\mathrm{planet}}$ denotes planetary mass.  $\dot{M}_{\mathrm{MRI,total}}$ is the total mass accretion rate in the inner MRI front. This Equation implies that the solid particles supplied to the inner MRI front during planet formation are immediately accreted by the planet as shown in Figure 6. From Equations~(\ref{eq:v_r_planet}) and (\ref{eq:gamma_0}) in the main text and Equation~(7) in EI. Equation~(\ref{eq:planet_accretion_rate}) can be derived as:
\begin{eqnarray}
\frac{dM_{\mathrm{planet}}}{dr_{\mathrm{planet}}}
&\simeq&
\frac{
\dot{M}_{\mathrm{MRI,total}} \, c_{\mathrm{s}}^2 \, M_*^{1/2}
}{
2 \, \Sigma \, G^{3/2} \, M_{\mathrm{planet}} \, r_{\mathrm{planet}}^{1/2}
}
\left( \frac{\Gamma}{\Gamma_0} \right)^{-1},
\label{eq:dMdr_planet_MRI}
\end{eqnarray}
where $\Gamma/\Gamma_0$ is normalized gravitational torque. By integration Equation~(\ref{eq:dMdr_planet_MRI}) over $M_{\mathrm{planet}}$ 
from $0$ to $M_{\mathrm{planet}}^*$ (the final planetary mass) and $r_{\mathrm{planet}}$ from $r_{\mathrm{in}}$  
(the semimajor axis of the inner MRI front) to $r_{\mathrm{in}} + \Delta R$
\begin{eqnarray}
M_{\mathrm{planet}}^{*2} &\simeq& 
\frac{
2 \dot{M}_{\mathrm{MRI,total}} c_{\mathrm{s}}^2 M_*^{1/2}
}{
\Sigma G^{3/2}
}
\left( \frac{\Gamma}{\Gamma_0} \right)^{-1} r_{\mathrm{in}}^{1/2}
\left\{ 
\left( \frac{r_{\mathrm{in}} + \Delta R}{r_{\mathrm{in}}} \right)^{1/2} - 1
\right\}.
\label{eq:mass_from_MRI_integral_a}
\end{eqnarray}
Here, the column density, $\Sigma$, and midplane temperature, $T_{\mathrm{m}}$, are assumed to remain almost constant from $r_{\mathrm{in}}$ to $r_{\mathrm{in}} + \Delta R$. $\Delta R$ 
is given as follows, representing a multiple of the Hill radius, denoted by $D$:
\begin{eqnarray}
\Delta R = D \left( \frac{M_{\mathrm{planet}}}{3 M_*} \right)^{1/3} r_{\mathrm{in}}.
\label{eq:width_growth_zone}
\end{eqnarray}
In the main text, $D=5$ and $r_{\mathrm{in}} = r_{\mathrm{planet}}$ were used in Equation~(\ref{eq:delta_R}). Since $r_{\mathrm{in}} \gg \Delta R$, Equation~(\ref{eq:mass_from_MRI_integral_a}) becomes as follow:
\begin{eqnarray}
M_{\mathrm{planet}}^{*2} &\simeq&
\frac{
\dot{M}_{\mathrm{MRI,total}} \, c_{\mathrm{s}}^2 \, M_*^{1/2} \, \Delta R
}{
\Sigma \, G^{3/2} \, r_{\mathrm{in}}^{1/2}
}
\left( \frac{\Gamma}{\Gamma_0} \right)^{-1}.
\label{eq:mass_from_MRI_integral_b}
\end{eqnarray}
From Equations~(\ref{eq:Sigma}), (\ref{eq:midplane_temperature}), (\ref{eq:width_growth_zone}), (\ref{eq:mass_from_MRI_integral_b}), and Equations~(6) and (7) for EI:
\begin{eqnarray}
M_{\mathrm{planet}}^{*5/3} &\simeq&
\frac{
D \, \dot{M}_{\mathrm{MRI,total}} \, c_{\mathrm{s}}^{2} \, M_*^{1/6} \, r_{\mathrm{in}}^{1/2}
}{
3^{1/3} \, \Sigma \, G^{3/2}
}
\left( \frac{\Gamma}{\Gamma_0} \right)^{-1} \nonumber \\
&=&
\left\{
\frac{
3^{10} \pi
}{
2^{28}
}
\frac{
D^{9} \, \dot{M}_{\mathrm{MRI,total}}^{9} \, \gamma^{9} \, \alpha_{\mathrm{act}}^{5} \, \beta_{\kappa}^{4} \, M_*^{3} \, k_{\mathrm{B}}^{14} \, T_{\mathrm{m}}
}{
\dot{M} \, \sigma^{4} \, (\mu m_{\mathrm{H}})^{14} \, G^{12}
}
\right\}^{1/9}
\left( \frac{\Gamma}{\Gamma_0} \right)^{-1}.
\label{eq:mass_from_MRI_integral_c}
\end{eqnarray}

This equation describes the relationship between the mass accretion rate at the inner MRI front and planetary mass.

\section{NORMALIZED GRAVITATIONAL TORQUE ($\Gamma/\Gamma_0$)}
\setcounter{equation}{0}
\renewcommand{\theequation}{G\arabic{equation}}

From Equations~(\ref{eq:Sigma}), (\ref{eq:midplane_temperature}), and Equations~(6), and (7) in EI:
\begin{eqnarray}
\Sigma &\simeq&
\left\{
\frac{
2^{28} \, \dot{M}^9 \, \sigma^4 \, (\mu m_{\mathrm{H}})^{13} \, (G M_*)^{5/2}
}{
3^{21} \, \pi^9 \, \gamma^{17} \, \alpha_{\mathrm{act}}^{13} \, \beta_{\kappa}^4 \, k_{\mathrm{B}}^{13}
}
\label{eq:Sigma_vs_r}
\right\}^{1/17}
r^{-15/34}.
\end{eqnarray}
Thus, from Equation~(\ref{eq:Sigma_vs_r}):
\begin{eqnarray}
\alpha &=& -\frac{\ln \Sigma}{\ln r} \ \simeq\ \frac{15}{34}.
\label{eq:alpha_slope_sigma}
\end{eqnarray}
From Equation~(\ref{eq:midplane_temperature}):
\begin{eqnarray}
\beta &=& -\frac{\ln T_{\mathrm{m}}}{\ln r} \ \simeq\ \frac{18}{17}.
\label{eq:beta_slope_Tm}
\end{eqnarray}
From Equation~(\ref{eq:Sigma}), and Equations~(6), (7), and (26) in EI, the optical depth ($\tau$) given by:
\begin{eqnarray}
\tau &=&
\left(
\frac{2^4}{3^4 \pi} 
\frac{
\dot{M} \, \beta_{\kappa}^2 \, \sigma \, (\mu m_{\mathrm{H}})^2 \, T_{\mathrm{m}}^{7/2}
}{
k_{\mathrm{B}}^2 \, \alpha_{\mathrm{act}}^2
}
\right)^{1/3} \nonumber \\[1em]
&\simeq&
\left(1.29 \times 10^3\right) 
\left( \frac{\dot{M}}{10^{-7.08} \, M_\odot \, \mathrm{yr}^{-1}} \right)^{1/3} 
\left( \frac{T_{\mathrm{m}}}{1000\,\mathrm{K}} \right)^{7/6}.
\end{eqnarray}
The isochoric heat capacity ($c_{\mathrm{V}}$) given by:
\begin{eqnarray}
c_{\mathrm{V}} &=&
\frac{2.5 \, k_{\mathrm{B}}}{\mu m_{\mathrm{H}}}.
\label{eq:cV}
\end{eqnarray}
Because the optical depth $\tau$ is assumed to be greater than $500$ by EI, Equation~(8b) in the text is approximated as follows:
\begin{eqnarray}
\Theta &=&
\frac{
c_{\mathrm{V}} \, \Sigma \, \Omega_{\mathrm{planet}}
}{
12 \pi \sigma T_{\mathrm{m}}^{3}
}
\left( \frac{3 \tau}{8} \right).
\label{eq:Theta}
\end{eqnarray}
From Equations~(\ref{eq:optical_depth}), (\ref{eq:kappa}), (\ref{eq:Sigma}), (\ref{eq:midplane_temperature}), (\ref{eq:cV}) and (\ref{eq:Theta}), and Equations~(5) and (6) in EI:
\begin{eqnarray}
\Theta &\simeq& 
\frac{2.5 \times 2}{3^3 \pi} \frac{1}{\gamma \alpha_{\mathrm{act}}} = 18.6.
\label{eq:Theta_value}
\end{eqnarray}
From Equations~(8a) in the main text and Equations~(\ref{eq:alpha_slope_sigma}), (\ref{eq:beta_slope_Tm}), and (\ref{eq:Theta_value}):
\begin{eqnarray}
\frac{\Gamma}{\Gamma_0} \simeq 1.21.
\end{eqnarray}

\section{MASS ACCRETION RATE TO THE INNER MRI FRONT}
\setcounter{equation}{0}
\renewcommand{\theequation}{H\arabic{equation}}

The mass accretion rate to the inner MRI front can be expressed as follows:
\begin{eqnarray}
\dot{M}_{\mathrm{MRI,total}} \simeq 2 \pi r \Sigma f_{\mathrm{p}} v_{\mathrm{mp}}.
\label{eq:Mdot_mri_total_a}
\end{eqnarray}
From Equations~(\ref{eq:Sigma}), (\ref{eq:midplane_temperature}), (\ref{eq:time_vzpp_equals_vB_b}), (\ref{eq:ln_mass_growth}), (\ref{eq:vmp}), and (\ref{eq:Mdot_mri_total_a}), Equations~(5) for EI becomes:
\begin{eqnarray}
\dot{M}_{\mathrm{MRI,total}} &\simeq&
\left(
\frac{\pi^{17}}{2^{17} \times 3^{22}}
\right)^{1/18}
\left\{
\ln \left(
\frac{m_{\mathrm{p},k}^*}{m_{0,k}^*}
\right) + \frac{2 \times 3}{5}
\right\}^{-1} 
\nonumber \\[1em]
&& \times
\left\{
\frac{
f_{\mathrm{p}}^{18} \dot{M}^{5} \sigma^{2} (\mu m_{\mathrm{H}})^7 (G M_*)^6
}{
\gamma^9 \alpha_{\mathrm{act}}^7 \beta_{\kappa}^2 k_{\mathrm{B}}^7 T_{\mathrm{m}}^{1/2}
}
\right\}^{1/9}.
\label{eq:Mdot_mri_total_b}
\end{eqnarray}

From Equations~(\ref{eq:mass_from_MRI_integral_c}) and (\ref{eq:Mdot_mri_total_b}):
\begin{eqnarray}
M_{\mathrm{planet}}^* &\simeq&
\left(
\frac{\pi^{19}}{2^{73} \times 3^{2}}
\right)^{1/30}
\left\{
\ln(10) \, \log_{10} \left(
\frac{m_{{\mathrm{p}},k}^*}{m_{0,k}^*}
\right) + \frac{2 \times 3}{5}
\right\}^{-3/5}
\nonumber \\[1em]
&& \times
\left\{
\frac{
D^{9} f_{\mathrm{p}}^{18} \dot{M}^{4} \beta_{\kappa}^{2} k_{\mathrm{B}}^{7} T_{\mathrm{m}}^{1/2} M_*^{9}
}{
\alpha_{\mathrm{act}}^{2} \sigma^{2} (\mu m_{\mathrm{H}})^{7} G^{6}
}
\right\}^{1/15}
\left( \frac{\Gamma}{\Gamma_0} \right)^{-3/5}.
\label{eq:Mplanet_star}
\end{eqnarray}

Using Equations~(\ref{eq:log_mass_ratio_scaling}) and (\ref{eq:Mplanet_star}):
\begin{eqnarray}
M_{\mathrm{planet}}^* &\simeq& 0.9 \left[
\frac{
\ln(10) \left[
3 \log_{10} \left( f_{\mathrm{p}} / (1.25 \times 10^{-2}) \right) + 8.4
\right] + (2 \times 3) / 5
}{21}
\right]^{-3/5} \left( \frac{\Gamma}{1.21 \Gamma_0} \right)^{-3/5} \left( \frac{D}{5} \right)^{3/5} \nonumber \\
&& \times \left( \frac{T_{\mathrm{m}}}{1000\,\mathrm{K}} \right)^{1/30} \left( \frac{M_*}{M_\odot} \right)^{3/5} \left( \frac{\dot{M}}{10^{-7.08} M_\odot \mathrm{yr}^{-1}} \right)^{4/15} \left( \frac{f_{\mathrm{p}}}{1.25 \times 10^{-3}} \right)^{6/5} M_\oplus.
\end{eqnarray}
This equation corresponds to Equation~(\ref{eq:planetmass}) in the main text.

\section{TOTAL MASS ACCRETION TO THE INNER MRI FRONT FROM TERRESTRIAL PLANETS FORMATION REGION}
\setcounter{equation}{0}
\renewcommand{\theequation}{I\arabic{equation}}

The total mass of particles in a region, $M_{\mathrm{p,disk}}$, with orbital semi-major axis, $r$, and width, $\Delta r$:
\begin{eqnarray}
M_{\mathrm{p,disk}} &=&
2 \pi r \, \Delta r \, f_{\mathrm{p}} \, \Sigma.
\label{eq:Mp_disk}
\end{eqnarray}
Because the total mass is supplied to the MRI front each time, it can be rewritten as follows:
\begin{eqnarray}
M_{\mathrm{p,disk}} &=&
\Delta M_{\mathrm{Rocky}}.
\label{eq:Mp_disk_dMrocky}
\end{eqnarray}
From Equation~(\ref{eq:v_r_planet}) in the main text:
\begin{eqnarray}
\frac{dr}{dT_{\mathrm{m}}} &\simeq& -\frac{17}{18} \frac{r}{T_{\mathrm{m}}}.
\label{eq:dr_dTm}
\end{eqnarray}
Using Equations~(\ref{eq:Sigma}), (\ref{eq:midplane_temperature}), (\ref{eq:Mp_disk}), (\ref{eq:Mp_disk_dMrocky}), (\ref{eq:dr_dTm}), and Equations~(6) and (7) in EI:
\begin{eqnarray}
\frac{dM_{\mathrm{Rocky}}}{dT_{\mathrm{m}}} &\simeq& 
- \left\{
\frac{
17^9
}{
2^7 \times 3^{26} \times \pi^2
}
\frac{
f_{\mathrm{p}}^9 \, \beta_{\kappa} \, \dot{M}^{11} \, (\mu m_{\mathrm{H}})^{10} \, (G M_*)^6
}{
\gamma^9 \, \alpha_{\mathrm{act}}^{10} \, k_{\mathrm{B}}^{10} \, \sigma \, T_{\mathrm{m}}^{89/4}
}
\right\}^{1/9}.
\label{eq:dMrocky_dTm}
\end{eqnarray}
By integrating Equation~(\ref{eq:dMrocky_dTm}) over $M_{\mathrm{Rocky}}$ from $0$ to $M_{\mathrm{Rocky}}$ and $T_{\mathrm{m}}$ from $1000$ to $180\,\mathrm{K}$, we obtain:
\begin{eqnarray}
M_{\mathrm{Rocky}} &\simeq&
\left(
180^{-53/36} - 1000^{-53/36}
\right)
\left[
\frac{2^{11} \times 17^9}{3^8 \times 53^9 \times \pi^2}
\frac{
f_{\mathrm{p}}^9 \, \beta_{\kappa} \, \dot{M}^{11} \, (\mu m_{\mathrm{H}})^{10} \, (G M_*)^6
}{
k_{\mathrm{B}}^{10} \, \sigma \, \gamma^9 \, \alpha_{\mathrm{act}}^{10}
}
\right]^{1/9}
\nonumber \\[1em]
&\simeq&
1.7
\left( \frac{M_*}{1 M_\odot} \right)^{2/3}
\left( \frac{\dot{M}}{10^{-7.08} \, M_\odot \, \mathrm{yr}^{-1}} \right)^{11/9}
\left( \frac{f_p}{1.25 \times 10^{-3}} \right)
M_\oplus.
\label{eq:Mrocky}
\end{eqnarray}

From Equations~(\ref{eq:log_mass_ratio_scaling}), (\ref{eq:Mplanet_star}) and (\ref{eq:Mrocky}):
\begin{eqnarray}
\frac{M_{\mathrm{Rocky}}}{M_{\mathrm{planet}}^*} &\simeq&
\left(180^{-53/36} - 1000^{-53/36}\right)
\left(
\frac{2^{329} \times 17^{90}}{3^{74} \times 53^{90} \times \pi^{77}}
\right)^{1/90}
\left\{
\ln(10) \log_{10} \left( \frac{m_{\mathrm{p},k}^*}{m_{0,k}^*} \right) + \frac{2 \times 3}{5}
\right\}^{3/5}
\nonumber \\[1em]
&& \times
\left(
\frac{
\dot{M}^{43} \, \sigma \, (\mu m_{\mathrm{H}})^{71} \, G^{48} \, M_*^3
}{
D^{27} \, f_{\mathrm{p}}^9 \, k_{\mathrm{B}}^{71} \, T_{\mathrm{m}}^{3/2} \, \beta_{\kappa} \, \gamma^{45} \, \alpha_{\mathrm{act}}^{44}
}
\right)^{1/45}
\left( \frac{\Gamma}{\Gamma_0} \right)^{3/5}
\nonumber \\[1em]
&\simeq&
1.9
\left[
\frac{
3 \ln(10) \log_{10} \left\{ f_{\mathrm{p}} / (1.25 \times 10^{-2}) \right\} + 8.4 \ln(10) + (2 \times 3) / 5
}{
21
}
\right]^{3/5}
\left( \frac{\Gamma}{1.21 \Gamma_0} \right)^{3/5}
\left( \frac{D}{5} \right)^{-3/5}
\nonumber \\[1em]
&& \times
\left( \frac{T_{\mathrm{m}}}{1000\,\mathrm{K}} \right)^{-1/30}
\left( \frac{M_*}{1 M_\odot} \right)^{1/15}
\left( \frac{\dot{M}}{10^{-7.08} \, M_\odot \, \mathrm{yr}^{-1}} \right)^{43/45}
\left( \frac{f_{\mathrm{p}}}{2.5 \times 10^{-3}} \right)^{-1/5}.
\label{eq:Mrocky_ratio}
\end{eqnarray}
In the case of $\dot{M} = 10^{-7.08} \, M_\odot \, \mathrm{yr}^{-1}$, Equation~(\ref{eq:Mrocky_ratio}) suggests that approximately two 
Earth-like planets could form, which may be interpreted as the formation of Earth and Venus.

\bibliographystyle{aasjournal}
\bibliography{Nimura_and_Ebisuzaki_2025}

\end{document}